\documentclass[structabstract]{aa}
\pdfoutput=1

\usepackage{natbib}
\usepackage{txfonts}
\usepackage{graphicx}
\usepackage{t1enc}
\usepackage{epsfig}
\usepackage{epstopdf}
\usepackage{color}

\bibpunct{(}{)}{;}{a}{}{,} 

\newcommand{\mum}   {~$\mu$m}
\newcommand{\kms}   {~km~s$^{-1}$}

\newcommand{\cmt}   {~cm$^{-3}$}
    
\newcommand{\lo}    {$L_{\odot}$}

\newcommand{\nh}    {NH$_3$}
\newcommand{\eg}    {e.\,g.}
\newcommand{\ie}    {i.\,e.}
\newcommand{\trot}  {$T_\mathrm{rot}$}
\newcommand{\tk}    {$T_\mathrm{k}$}
\newcommand{\texc}  {$T_\mathrm{exc}$}
\newcommand{\tbg}   {$T_{\mathrm{bg}}$}
\newcommand{\g}     {G79.29+0.46}

\begin{document}

\title{Ammonia observations in the LBV nebula G79.29+0.46}
\subtitle{Discovery of a cold ring and some warm spots}

\author{J. R. Rizzo\inst{1}
  \and
  Aina Palau\inst{2}
  \and 
  F. Jim\'enez-Esteban \inst{1,3}
  \and
  C. Henkel\inst{4,5}
}

\institute{Centro de Astrobiolog\'{\i}a (INTA-CSIC), Ctra. M-108, km.~4, 
           E-28850 Torrej\'on de Ardoz, Madrid, Spain\\
  \email{ricardo@cab.inta-csic.es}
  \and
  Institut de Ci\`encies de l'Espai (CSIC-IEEC), Campus UAB Facultat de 
  Ci\`encies, Torre C5-parell 2, E-08193 Bellaterra, Spain
  \and
  Suffolk University, Madrid Campus, 
  C/ Valle de la Vi\~na 3, E-28003 Madrid, Spain
  \and
  Max-Planck-Institut f\"ur Radioastronomie, Auf dem 
  H\"ugel 69, D-53121 Bonn, Germany
  \and
  Astronomy Department, King Abdulaziz University, P.O. Box 80203,
  Jeddah 21589, Saudi Arabia 
}

\date{Received / Accepted}

\authorrunning{Rizzo et al.}
\titlerunning{\nh\ associated with G79.29+0.46}

\abstract
    {The surroundings of Luminous Blue Variable (LBV) stars are excellent 
    laboratories to study the effects of their high UV radiation, powerful 
    winds, and strong ejection events onto the surrounding gas and dust.}
    {We aim at determining the physical parameters of the dense gas near \g, an 
    LBV-candidate located at the centre of two concentric infrared rings which 
    may interact with the infrared dark cloud (IRDC) G79.3+0.3.}
    {The Effelsberg 100\,m telescope was used to observe the \nh (1,1), (2,2) 
    emission in a field of view of $7'\times7'$ including the infrared rings 
    and a part of the IRDC. In addition, we observed particular positions in 
    the \nh(3,3) transition toward the strongest region of the IRDC, which is 
    also closest to the ring nebula.}
    {
    We report here the first coherent ring-like structure of dense \nh\ gas 
    associated with an evolved massive star. It is well traced in both ammonia 
    lines, surrounding an already known infrared ring nebula; its column 
    density is two orders of magnitude lower than the IRDC. The \nh\ emission 
    in the IRDC is characterized by a low and uniform rotational temperature 
    (\trot\ $\sim10$\,K) and moderately high opacities in the (1,1) line. The 
    rest of the observed field is spotted by warm or hot zones 
    (\trot\ $>30$\,K) and characterized by optically thin emission of the (1,1) 
    line. The \nh\ abundances are about 10$^{-8}$ in the IRDC, and 
    10$^{-10}$--10$^{-9}$ elsewhere. The warm temperatures and low abundances 
    of \nh\ in the ring suggest that the gas is being heated and 
    photo-dissociated by the intense UV field of the LBV star. An outstanding 
    region is found to the south-west (SW) of the LBV star within the IRDC. The 
    \nh\ (3,3) emission at the centre of the SW region reveals two velocity 
    components tracing gas at temperatures $>30$~K. Of particular interest is 
    the northern edge of the SW region, which coincides with the border of the 
    ring nebula and a region of strong 6~cm continuum emission; here, the 
    opacity of the (1,1) line and the \nh\ abundance do not decrease as 
    expected in a typical clump of an isolated cold dark cloud. This strongly 
    suggests some kind of interaction between the ring nebula (powered by the 
    LBV star) and the IRDC. We finally discuss the possibility of \nh\ 
    evaporation from the dust grain mantles due to the already known presence 
    of low-velocity shocks in the area.}
    {The detection of the \nh\ associated with this LBV ring nebula, as well as 
    the special characteristics of the northern border of the SW region, 
    confirm that the surroundings of \g\ constitutes an exemplary scenario, 
    which is worth to be studied in detail by other molecular tracers and 
    higher angular resolutions.}
    

\maketitle

\section{Introduction}

Massive stars are probably the most important sources providing thermal and 
mechanical energy to the interstellar medium (ISM). They are few in number and 
remain on the Main Sequence during some million years only. One of the shortest 
(up to a few $10^4$ yr) and most spectacular stages of their evolution is 
reached when they become luminous blue variable (LBV) stars. With bolometric 
luminosities $>3\times10^5$~\lo, LBV stars are characterized by their heavy 
mass loss and both photometric and spectroscopic variability, and are supposed 
to be the progenitors of Wolf-Rayet (WR) stars \citep{lan94,mae94} or, as 
suggested by other authors, even the direct progenitors of type II supernovae 
with dense circumstellar material \citep[\eg][]{gal07,smi07}. A key subject of 
current study about the LBV stars is the mechanism which drives the variability 
and episodic mass ejections. Since the winds and mass ejections leave their 
fingerprints on the surrounding material, the study of the distribution, 
composition and physical properties of the circumstellar gas and dust would put 
constraints on the stellar wind models \citep[\eg][]{gar95}.

This circumstellar material, at up to 1\,pc distance from the star, has been 
traditionally studied in the optical/IR, tracing emission nebulae consisting of 
ionized and neutral atomic gas and the (often) warm dust. However, the LBV 
environment is presumably also rich in molecules, (1) because the stars are 
active dust producers and (2) because there should be shocks, caused by the 
stellar wind, triggering a shock induced chemistry. In addition, since the 
atmospheric abundances of LBVs are expected to be He-enriched and near 
nuclear-CNO equilibrium, it is theoretically expected that the ejecta from LBVs 
will be rich in nitrogenated species 
\citep[\eg][and references therein]{smi98}. 

A clear-cut case of an environment rich in molecules is \object{NGC\,2359}. In 
this WR nebula, \citet{riz01a} have detected the \nh\ (1,1) and (2,2) 
metastable lines, which was the first detection of a polyatomic molecule in the 
surroundings of an evolved massive star. \nh\ was detected towards a peak of 
the CO emission \citep{riz01b,riz03a}, in coincidence with other complex 
molecules \citep[such as HCO$^+$, CS, CN, and HCN; see][]{riz03b}. These 
findings revealed the presence of dense and warm molecular gas, which can be 
traced by N-bearing molecules. The high ammonia abundance and kinetic 
temperature derived ($10^ {-8}$ and 30\,K, respectively) are interpreted in 
terms of a shock-induced chemistry. The same molecules detected in 
\object{NGC\,2359} (and some isotopomers) were later reported in the 
Homunculus, the circumstellar nebula around the well known LBV $\eta$\,Carina 
\citep{smi06,loi12}.

Among the N-bearing molecules usually studied, \nh\ is particularly interesting 
due to its relatively high abundance and ubiquity. Moreover, its properties 
allow an accurate determination of the kinetic temperature, avoiding usual 
observational issues like calibration and pointing \citep{ho83}. The detection 
of \nh\ in the environment of NGC\,2359 and $\eta$\,Carina encourages studies 
of additional LBV star environments. 

The nebula \object{\g} constitutes an excellent testbed to pursue this kind of 
studies. The combined effect of the high UV field, the steady powerful winds 
and the violent mass ejections have produced conspicuous structures in the 
nebula and beyond. \g, firstly discovered by its free-free emission 
\citep{hig94}, is a ring-like nebula formed and excited by a LBV candidate, 
surrounded by an outstanding pair of concentric rings seen in the infrared up 
to 24\mum\ \citep[see Fig.~1 of][]{jim10}, which is likely formed by material 
ejected by the central object in at least two different mass-loss episodes. The 
detection of a CO structure around the nebula, as observed in the 
$J=2\rightarrow1$ and $J=3\rightarrow2$ lines \citep{riz08}, demonstrated the 
existence of warm and/or high-density molecular gas. Immersed within the CO 
structure, there is a higher density clump which is claimed to be associated 
with the south-western side of the infrared nebula. This work also provided 
evidences for a low-velocity shock with velocities around 14--15\kms\ toward 
this CO clump. Subsequent CO $J=4\rightarrow3$ emission \citep{jim10} revealed 
two concentric slabs in the same region of the clump, filling the space between 
the two shells detected at 24\,$\mu$m, and in coincidence with a brightening of 
the 6~cm continuum emission of the nebula \citep{uma11}.

\g\ is part of the very complex \object{Cygnus X} region. It is projected in 
between two star-forming regions: the {\sc Hii} region \object{DR\,15} to the 
south-east \citep[see Fig. 11 of][]{kra10}, and the infrared dark cloud 
\object{IRDC\,G79.3+0.3} (hereafter the IRDC) towards the southern and western 
borders \citep{red03}. The local arm is seen tangentially in this direction 
\citep{dam85, ryg12}, which makes it difficult to determine distances and to 
identify mutual associations among targets associated with the Cyg X region. An 
update of the information currently available and a brief discussion of the 
distance is presented below, in Sect.~5.1. 

In this paper, we present \nh (1,1) and (2,2) maps observed with the Effelsberg 
100-m radio telescope towards the LBV nebula G79.29+0.46 and the nearby IRDC. 
In Sect.~2 we describe the observations. In Sect.~3 we present the structure of 
ammonia emission in different velocity ranges, revealing a ring-like structure 
associated with the previously known infrared shell, as well as spectra of \nh\ 
(3,3) towards two positions in the south-western region. In Section 4 we 
analyse the ammonia data in the field, and study the properties of the dense 
gas associated with the LBV nebula and with the IRDC. Finally, in Sect.~5, we 
discuss the most relevant properties of the dense gas in the region, and the 
possible interaction of the LBV nebula with the IRDC. The main conclusions are 
summarized in Sect.~6.

\section{Observations}

We have used the Effelsberg 100-m radio telescope of the Max-Planck-Institut 
f\"ur Radioastronomie to map the distribution of the (1,1) and (2,2) inversion 
lines of \nh, in a $\sim7'\times7'$ field around \g. The observations have 
been carried out in March 2008.

The 1.3 cm HEMT receiver was tuned to 23.7086\,GHz, a frequency lying 
midway between the rest frequencies of the two observed lines. We have used 
the FFTS spectrometer, which provides a bandwidth of 100\,MHz and a 
frequency resolution of $\sim6$\,kHz (0.08\kms). This spectrometer allowed 
the simultaneous observation of the two lines, with excellent velocity 
resolution. The whole data processing was done using the CLASS 
software\footnote{CLASS is part of the GILDAS software, developed by IRAM. 
See {\tt http://www.iram.fr/IRAMFR/GILDAS}/.}.
After baseline subtraction and calibration, the spectra have been smoothed 
to a resolution of 0.24\kms\ for further analysis.

The telescope beam size (half power beam width, HPBW) at the observed 
frequency is $\sim 40$\arcsec, and the main beam efficiency is 0.52. We have 
used \object{NGC\,7027} for continuum calibration, assuming an absolute flux 
of 5.58\,Jy at 23.7\,GHz \citep{ott94}. We have also used \object{DR21} (close 
to the observed field) as a line calibrator. Pointing was regularly checked, 
typically once every hour, and is accurate to within 7\arcsec.

We have done all the observations in position switching mode, with the relative 
reference position at 200\arcsec\ in azimuth. Integration times per point 
varied between 5 and 10 minutes on-source, depending on weather conditions and 
elevations. Due to variable weather the system temperature varied between 30\,K 
and 94\,K, resulting in a {\it rms} noise (1-sigma) between 50\,mK and 350\,mK 
for a velocity spacing of 0.24\kms. All temperatures throughout this paper are 
on a main-beam scale $(T_\mathrm{MB})$. 

The whole field was initially sampled every 40\arcsec. A particular region of 
interest, corresponding to a multiple layered morphology in CO \citep{riz08} 
has been fully sampled (\ie~every 20\arcsec); this region is indicated in  
Figs.~1 to 4 by a square and is referred to as "the SW region" hereafter.

Furthermore, two deep-integration pointed observations of the (3,3) line were 
done towards specific positions. These positions correspond to \nh\ and CO 
peaks, as explained in Sect.~3.3. Identical technical conditions were applied 
in this case.

All the positions throughout this paper are offsets from the equatorial 
coordinates (J2000.0) of the exciting star, at (RA, Dec.) = 
$(20^\mathrm{h}31^\mathrm{m}42\fs29, +40\degr21'59\farcs1)$.

\begin{figure}
  \includegraphics[width=\columnwidth]{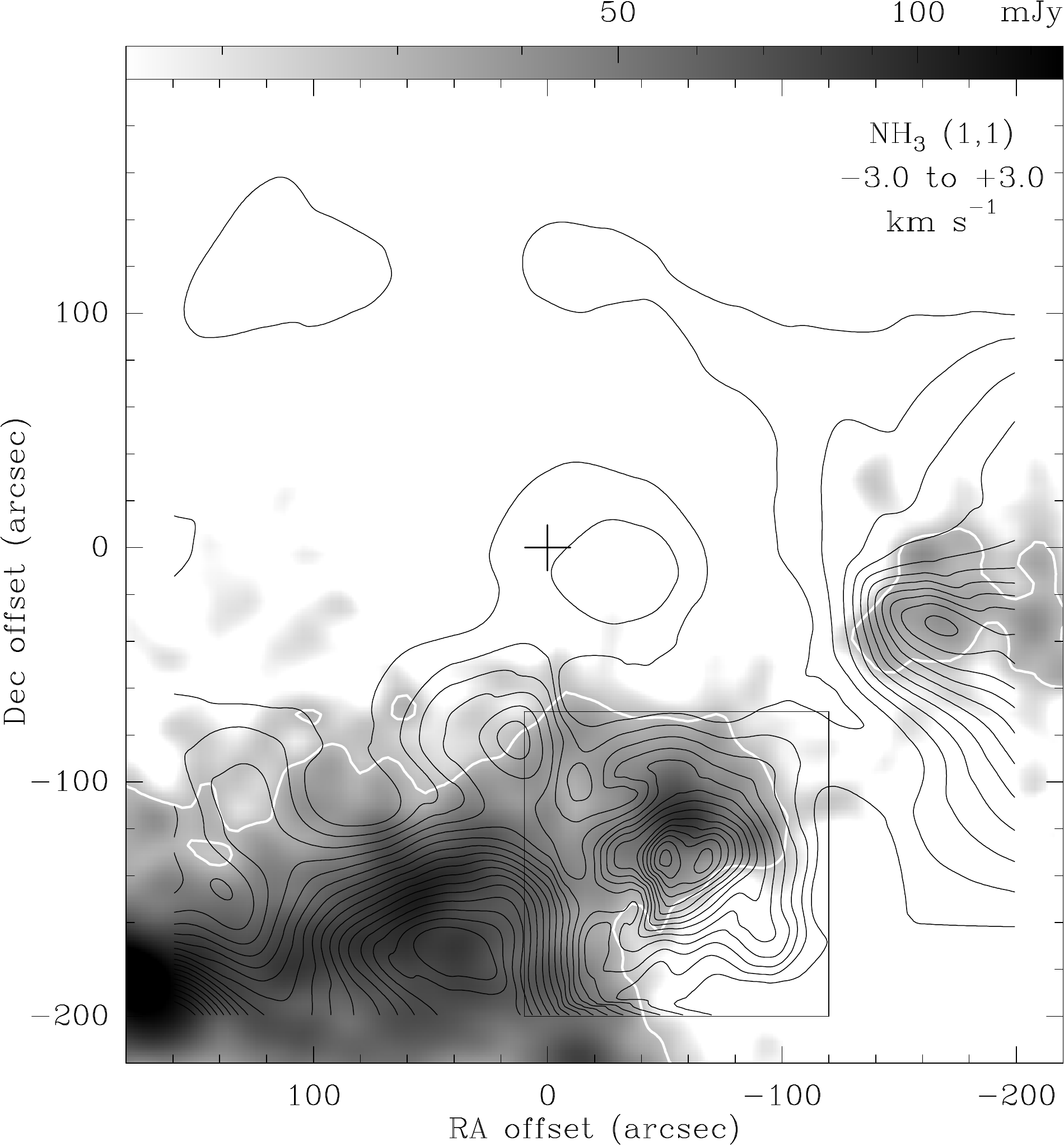}
  \caption{ \nh\,(1,1) line intensity map in direction to \g. Only the main 
    component has been integrated, in the velocity range indicated at the top 
    right corner. On grayscale, the 1.2\,mm continuum image of the MAMBO 
    cameras at the IRAM 30\,m telescope is also plotted \citep{mot07}. The 
    logarithmic intensity scale used is indicated in the top bar. \nh\ starting 
    contours and spacing are 0.2~K\kms. Equatorial coordinates are referred to 
    the star position (Sect.~2). The region was fully sampled within the square 
    indicated at the south west, and every 40\arcsec elsewhere. The white 
    contour, taken from the MAMBO data, is plotted to indicate the boundaries 
    of the IRDC. Most of the \nh\,(1,1) emission seems associated with the cold 
    dust traced by the 1.2\,mm continuum, although some low-level features are 
    also noted (see text).
    }
\end{figure}

\begin{figure}
  \includegraphics[width=\columnwidth]{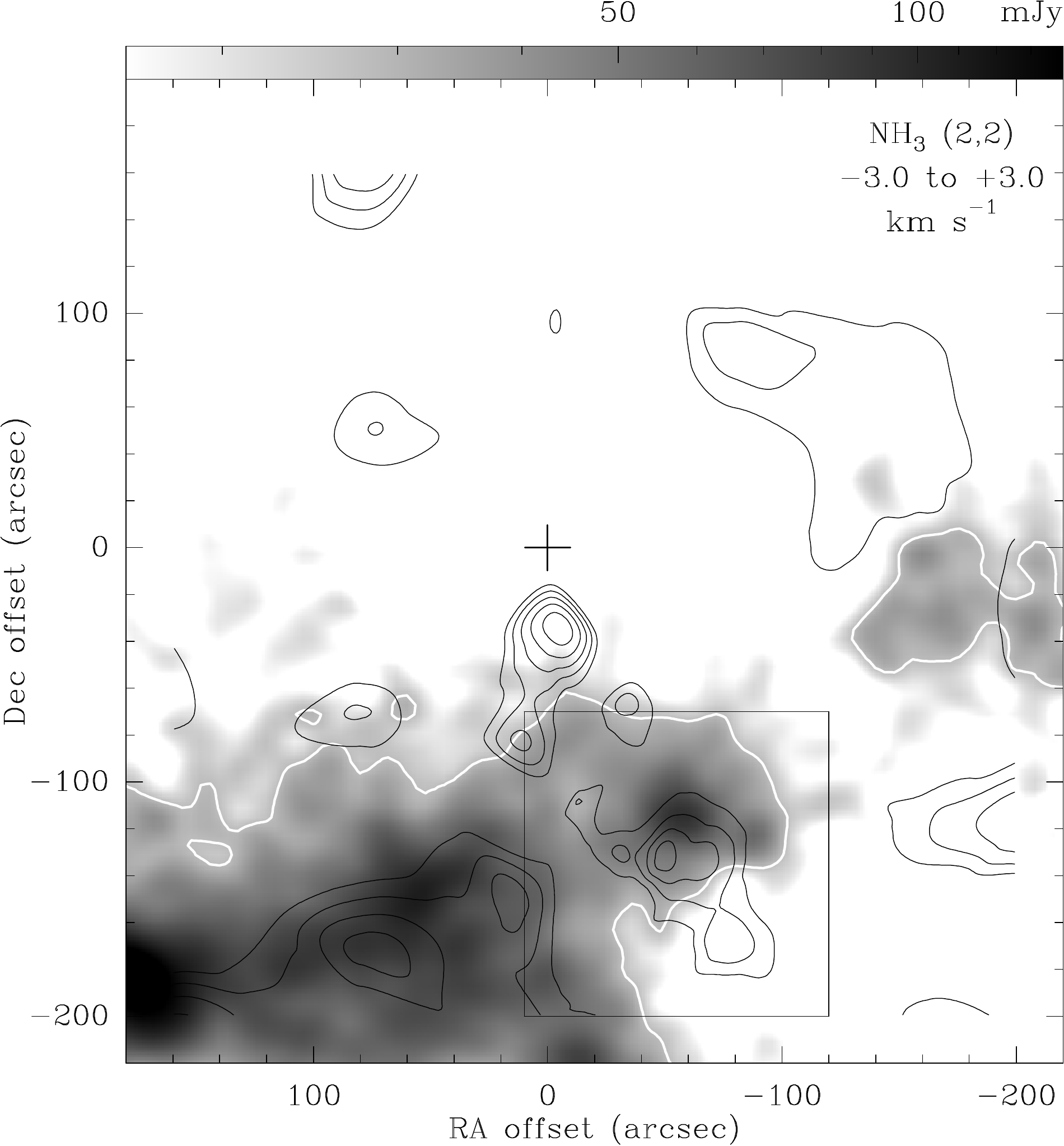}
  \caption{
    The same as Fig.~1 for the \nh\,(2,2) line. Contours start at 0.22~K\kms, 
    and are spaced by 0.11~K\kms. Although the most extended emission is well 
    correlated to the dark cloud in the south and west, the peak of the (2,2) 
    line arises from the $(0\arcsec, -40\arcsec)$ position and is not 
    correlated to the 1.2\,mm emission.
  }
\end{figure}

\section{Results}

\subsection{Overall emission}

The whole velocity range of \nh\ emission, from $-3$\kms\ to $+3$\kms, falls 
within the corresponding to Cyg~X and the Great Rift regions 
\citep{sch06,got12}. Channel maps of the emission of the (1,1) and (2,2) lines 
are presented in an online appendix, and the spectra are also accessible for 
download at CDS. The velocity-integrated maps of both lines, in the interval 
$(-3, +3)$\kms, are shown in Figs.~1 and 2. In Fig.~1, the (1,1) contours are 
superimposed on the 1.2\,mm continuum image from the MAMBO and MAMBO-2 cameras 
at the IRAM 30\,m telescope \citep{mot07}, smoothed to an angular resolution of 
$\sim$26\arcsec. The figure shows that the most intense \nh\ and 1.2~mm 
emission arises from the IRDC, extending from the south-east to the west of our 
field of view, and consisting of four main clumps. The peaks of ammonia do not 
agree with those of the 1.2~mm continuum. The SW region (the area inside the 
square) is the closest in projection to the ring of ionized nebula 
\citep{hig94,uma11}. It is worth noting that while the southern clump and the 
SW region are physically connected, the connection between the SW region and 
the western clump suffers a discontinuity both in dust and gas, even taking 
into account that their overall structure suggests they are part of the same 
large-scale cloud. There is also some weak emission, traced by the low-level 
contours; this low-level emission is particularly visible at the northern and 
western part of the field, and also close to its centre, near the LBV star.

The \nh\ (2,2) line emission shares some of the (1,1) features. Fig.~2 plots 
the \nh\ (2,2) emission in the same velocity range as in Fig.~1. Again, the 
most extended emission appears correlated to the 1.2\,mm continuum, and 
consequently to the IRDC; furthermore, some emission at the centre and the west 
is also present. It is notable that the peak of the whole map arises from a 
position close to the LBV star, at $(0\arcsec, -40\arcsec)$. The central part 
of the field is devoid of \nh\ (2,2) emission, except at the 
$(0\arcsec, -40\arcsec)$ position.

\begin{figure}
  \includegraphics[width=\columnwidth]{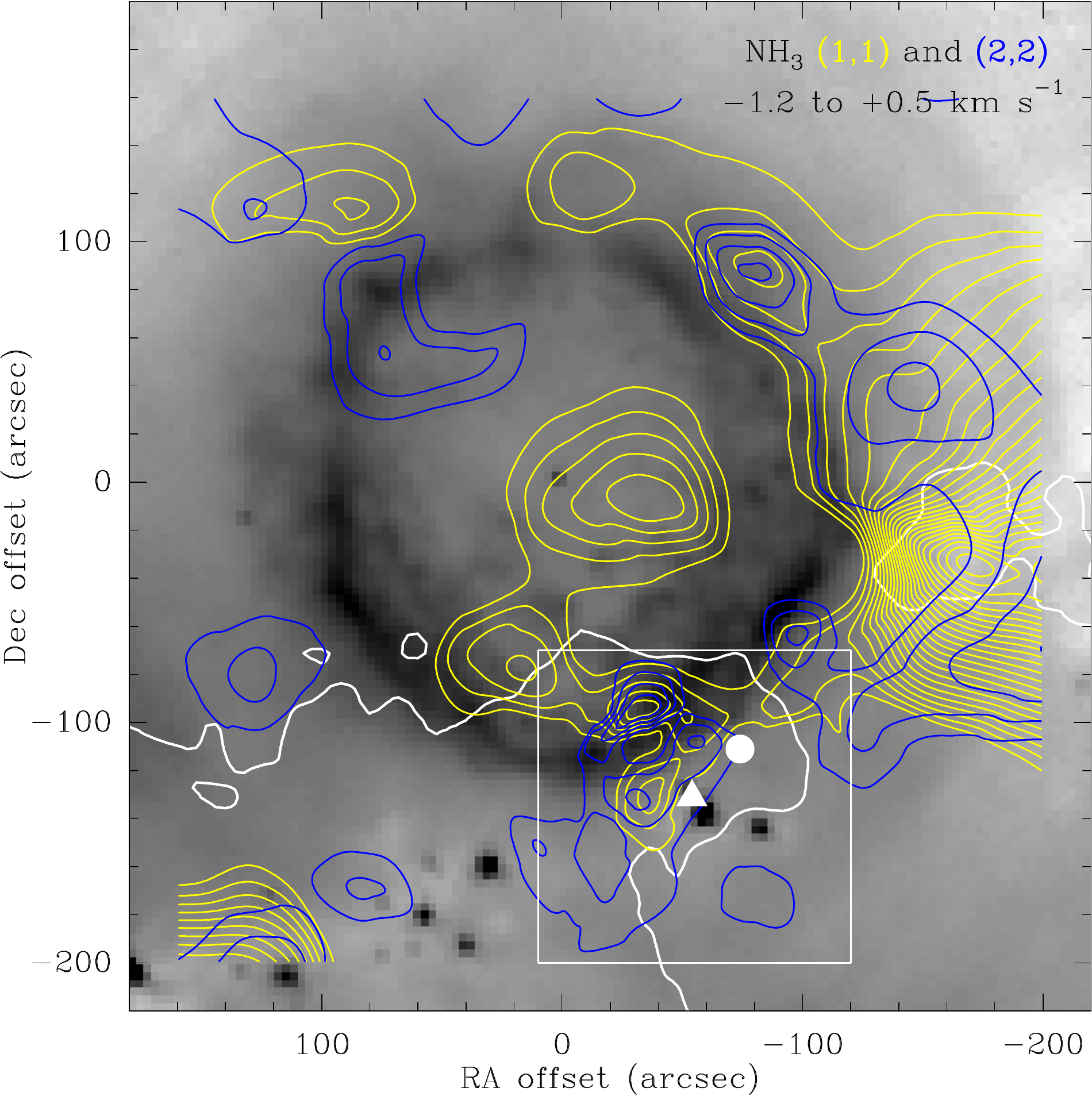}
  \caption{
    \nh\ (1,1) and (2,2) integrated line emission in the velocity range 
    $(-1.2, +0.5)$\kms, overlaid on the 70\mum\ continuum of Herschel/PACS.
    The white contour is the same depicted in Figs.~1 and 2. The fully sample 
    area is indicated by the white square. The white triangle (circle) points 
    to the position of the \nh\ (CO) peak. \nh(1,1) starting and spacing 
    contours (yellow) are 0.12 and 0.06~K\kms, respectively. \nh(2,2) starting 
    and spacing contours (blue) are 0.04~K\kms. In this velocity range, the 
    ammonia emission is dominated by a ring-like structure towards the outer 
    part of the ring nebula (see text).
  }
\end{figure}

\begin{figure}
  \includegraphics[width=\columnwidth]{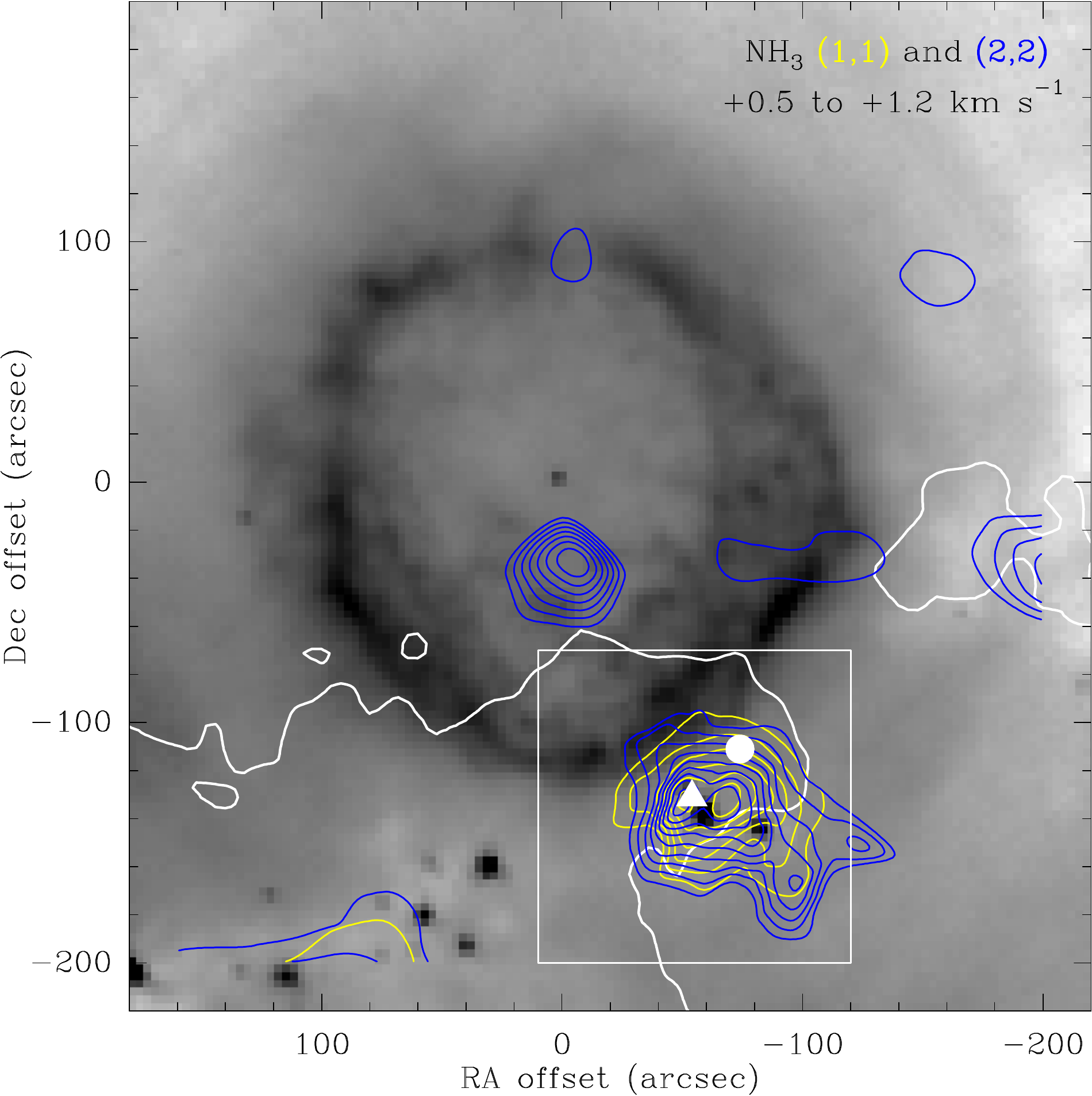}
  \caption{
    The same as Fig.~3, but for the the velocity range $(+0.5, +1.2)$\kms. 
    Starting contours and spacing of the \nh\ (1,1) line (yellow) are 
    0.6~K\kms\ and 0.2~K\kms, respectively. For the \nh\ (2,2) line, the 
    starting contours and spacing (blue) are 0.09 and 0.03~K\kms, 
    respectively. At this velocity range, the ammonia emission is particularly 
    intense towards the SW region (e.g., within the white square), 
    in coincidence with the most prominent CO emission. Another outstanding 
    feature is the (2,2) spot around $(0\arcsec, -40\arcsec)$, which has no 
    (1,1) counterpart.
  }
\end{figure}

\subsection{\nh\ associated with the LBV nebula}

In order to identify and further characterize the ammonia associated
with \g, we have thoroughly studied the \nh\ emission in different
velocity bins. After this analysis, two velocity ranges are particularly 
interesting. In Figs.~3 and 4, the emission of the (1,1) and (2,2) lines 
are sketched in such velocity intervals. 

Figure 3 shows the ammonia emission in the velocity range $(-1.2, +0.5)$\kms, 
overlaid to a Herschel/PACS image at 70\mum\,\footnote
{The Level-2 PACS image \citep{pog10} was obtained from 
the Herschel public archive (ObsID: 1342196767).}. 
An outstanding morphological correlation with the ring nebula (especially in 
the (1,1) line) is noted in this figure, where a large fraction of the 70\mum\ 
ring is surrounded by an ammonia feature. This \nh\ ring-like structure bounds 
the 70\mum\ ring to the north and west, but is projected above it towards the 
south. The (2,2) line also shows a similar morphology, although not as clear as 
the (1,1) line. The lowest-level contours of the (1,1) and (2,2) lines 
corresponds to a signal-to-noise (S/N) ratio greater than 4 and 1.5, 
respectively, which explains the lack of a better correlation in the (2,2) line. 

Moreover, the inner part of the nebula is also partially filled by (1,1) 
emission, without a (2,2) counterpart. Due to the undersampling and the limited 
S/N ratio, it is not possible to determine whether a physical connection 
between the ammonia ring and the emission at the field centre exists. If so, 
most of the \nh\ associated with the LBV nebula could be the surviving of a 
formerly complete \nh\ shell.

Figure 4 is similar to Fig.~3, but covers the second velocity range of 
interest: $(+0.5, +1.2)$\kms. Here, the \nh\ emission of both lines is 
concentrated towards the SW region, approximately at the centre of the fully 
sampled area. The (2,2) line emission also depicts a strong spot around 
$(0\arcsec, -40\arcsec)$ without a counterpart in the (1,1) line.

The velocity of the \nh\ ring does not match with the values reported for the 
CO shell \citep[see Figs.~6 and 7 of][]{jim10}, which are around $-3$\kms\ and 
$+3$\kms. This contradictory result is only apparent, because it accounts for 
intrinsic differences in the molecules under comparison. CO is very abundant 
and ubiquitous, optically thick and with a low critical density; therefore, it 
is expected the emission of CO in environments with a wider range of physical 
conditions than \nh. The velocity of the \nh\ ring fits well within the whole 
range of emission associated to the CO shell 
\citep[$-5$ to $+5$\kms; see][]{riz08}, but is close to the velocity of the 
local arm in this direction, and also the velocity of Cygnus X and the Great 
Cygnus Rift. The velocity ranges of the CO slabs reported by \citet{jim10} are 
merely those where the confusion effects are minor and a clear correlation with 
the infrared shell is noted.

In projection, the \nh\ ring bounds the ring nebula and the CO slabs, although 
the presence of any positional differences between CO and \nh\ should take into 
consideration the arguments presented above, and the fact that the comparison 
is made at different velocities. As we discuss below, the lack of ammonia at 
the ring nebula positions is consistent with the destruction of this molecule 
by the high UV field.

\subsection{\nh\ (3,3) in the SW region}

We devoted a special attention to the SW region, because it hosts a multiple 
layered CO structure \citep{riz08,jim10}, and is also claimed as a site of 
interaction between the LBV nebula and its environs \citep{uma11}.

The CO \citep{riz08,jim10} and dust-continuum emission (Figs.~1 and 2) are very 
well correlated in this region. Ammonia also depicts a strong and compact 
emission in the SW region, although does not peak at the same position as CO 
and dust. Hereafter, we will refer to the peaks of emission of ammonia and CO 
as the ``\nh\ peak'' and the ``CO peak''. The \nh\ peak is near the centre of 
the SW region, at $\sim(-54\arcsec, -131\arcsec)$, while the CO peak is shifted 
with respect to the \nh\ peak approximately by $(-20\arcsec, +20\arcsec)$. In 
order to better analyse these two points, we have done deep integrations of the 
(3,3) line.

\begin{figure}
  \centering
  \includegraphics[width=0.9\columnwidth]{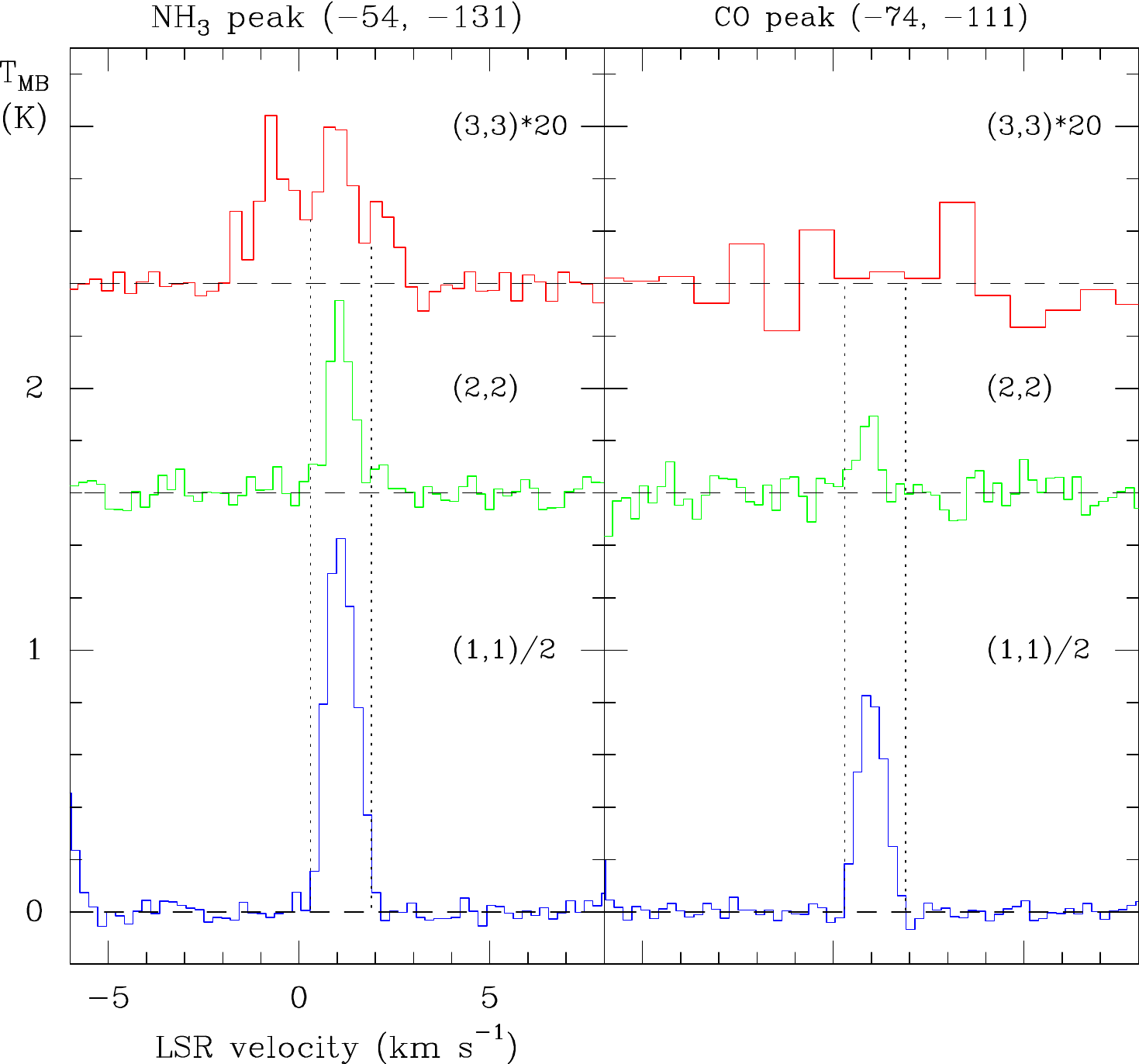}
  \caption{
    \nh\,(3,3) observations in the south-western part of the field. Two 
    positions (the \nh\ and the CO peaks) have been observed, indicated above 
    the boxes. The (1,1) and (2,2) spectra towards the same positions are also 
    plotted for comparison. Note that the (1,1) intensity scale is reduced by a 
    factor of two, and the (3,3) spectra are multiplied by 20. The (3,3) 
    spectrum towards the CO peak has been smoothed to 0.7\kms. Dashed vertical 
    lines are plotted at $+0.3$ and $+1.9$\kms\ to facilitate further 
    comparisons among the lines, and to separate the three components present 
    in the (3,3) line towards the \nh\ peak.
  }
\end{figure}

Both (3,3) spectra are presented in Fig.~5, together with the corresponding 
(1,1) and (2,2) lines. Dashed velocity marks have been added at $+0.3$ and 
$+1.9$ \kms\ to facilitate the comparisons.

The first feature in the figure is the lack of (3,3) emission in the CO peak. 
The ratio of the areas between the \nh\ peak and the CO peak --computed in the 
velocity range $(+0.3, +1.9)$\kms of the (3,3) line-- is greater than 16. On 
the other hand, the same ratios computed for the (1,1) and (2,2) lines are 1.8 
and 2.8, respectively. Therefore, the absence of emission in the (3,3) towards 
the CO peak is statistically significant.

A second feature in Fig.~5 is the presence of three velocity components in the 
(3,3) line towards the \nh\ peak. The velocity of the central component, 
roughly between the velocity marks, agrees with those of the (1,1) and (2,2) 
lines. However, the lower (blue) and higher (red) velocity components of the 
(3,3) line do not have any counterpart in the (1,1) or (2,2) line emission. As 
we analyse below, the blue and red components are characterized by higher 
temperatures than the central component.

\section{Analysis of the \nh\ emission}
\subsection{Derivation of physical parameters}

Ammonia is a rather ubiquitous molecule. Heavily destroyed in the presence of 
UV fields \citep[\eg][and references therein]{fue90}, it is present in cold 
dark clouds \citep{jij99,sep11} and also in regions affected by shocks 
\citep{taf95,zha02,pal07}. Since the very beginnings of molecular spectroscopy, 
it has been largely recognized as an excellent thermometer of the ISM 
\citep{ho83,gue85,mar09}.

From the joint analysis of the (1,1) and (2,2) metastable lines, we have 
derived several physical properties of the \nh\ gas. The details of the 
formulation are explained in an online appendix. This analysis was done at each 
position with significant emission ($>3\sigma$) in any of the two lines.

The (1,1) line emission arising from the IRDC is mostly optically thick, and 
the opacities were estimated through a hyperfine (hf) fitting. No satellite 
lines were detected elsewhere, and only a Gaussian fitting were done. All the 
positions with significant emission in the (2,2) line were also fitted by 
Gaussian profiles.

To estimate the ammonia abundance, $X($\nh$)$, we have used the 1.2 mm map of 
\citet{mot07} and computed the H$_2$ column density following \citet{mot98}. 
Assumptions and formulation are also exposed in the Appendix.

In the \nh\ peak of the SW region, where the (3,3) line was observed and 
detected, we proceeded with a Boltzmann diagram approach to determine the 
rotational temperature and column densities of the three velocity components. 
The procedure is also described in the Appendix.

\subsection{Rotational and kinetic temperatures}

The distribution of the rotational temperatures (\trot) is sketched in Fig.~6. 
The values are written at their corresponding positions, superimposed to a 
Spitzer image at 8\mum. 

The blue labels correspond to the points where the hf fitting was possible. All 
these positions belong to the IRDC (as said above), and are characterized by a 
very low and uniform \trot, in the range $7-11$\,K. The points having two 
values of \trot\ correspond to those cases where two velocity components were 
fitted.

The yellow labels correspond to those points with both lines fitted by Gaussian 
profiles, and with derived values of \trot\ lower than 40\,K. In case of 
non-detection of the (2,2) line, the values are expressed as upper limits. None 
of these points are correlated to the IRDC, and it is hard to establish a 
tendency across the observed field. 

The red labels correspond to those positions where the (1,1) line remains 
undetected, and therefore the values of \trot\ are lower limits. These ``hot 
spots'' are spread over the whole observed field, but again avoiding the IRDC 
positions. As the hot spots are located in the under-sampled area, we have to 
interpret them as small regions (less than $40\arcsec$) with special local 
conditions. The peak of the (2,2) line in Fig.~4 --offset $(0'',-40'')$-- has a 
lower limit of \trot\ of 63\,K.

\begin{figure}
  \includegraphics[width=\columnwidth]{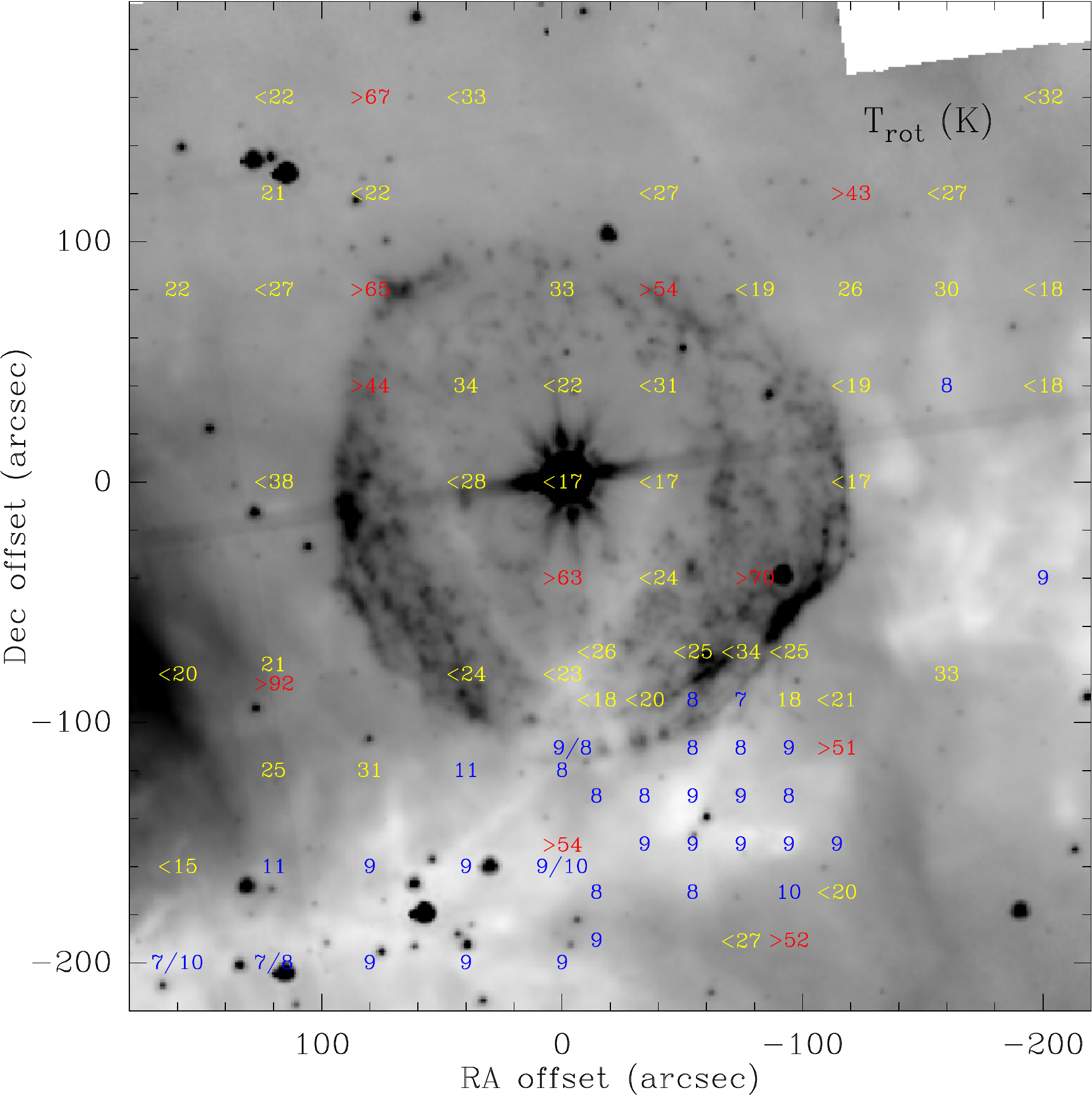}
  \caption{
    Distribution of the rotational temperatures, overlaid to the 8\mum\ image 
    of Spitzer. The values derived from hyperfine fitting of the (1,1) line 
    are labelled in blue; those derived from Gaussian fitting of both lines are 
    shown in yellow; red labels correspond to those positions having the highest 
    values of \trot\ (see text). In all cases, fitting of the (2,2) has been 
    obtained with a single Gaussian component. The low temperatures associated 
    with the IRDC, and higher values elsewhere are clearly revealed.
  }
\end{figure}

\begin{figure}
  \includegraphics[width=\columnwidth]{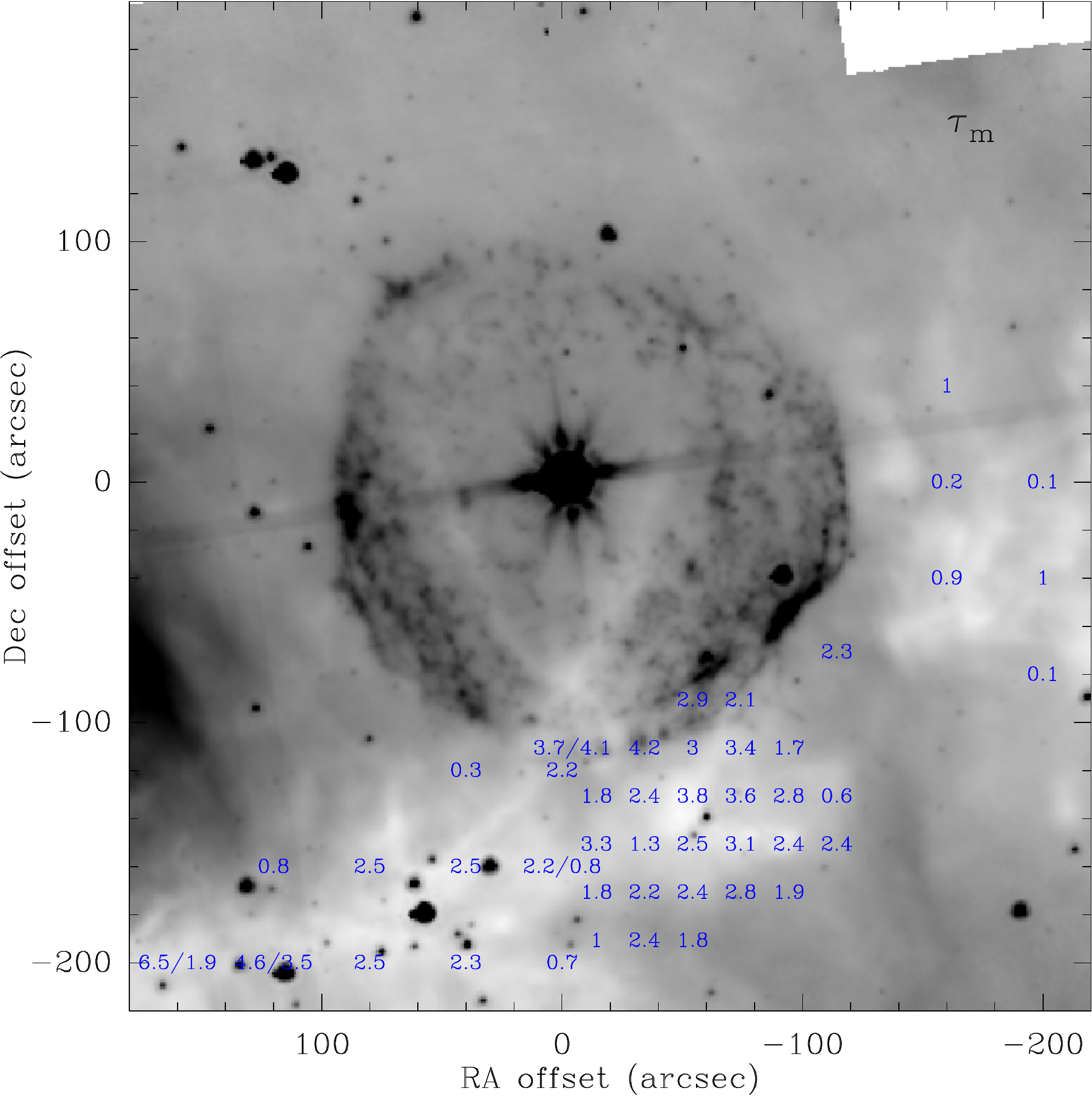}
  \caption{
    Distribution of the opacities, in all the positions where a hyperfine fitting 
    of the (1,1) line was possible, \ie\ mostly in the IRDC.
  }
\end{figure}

\begin{figure}
  \includegraphics[width=\columnwidth]{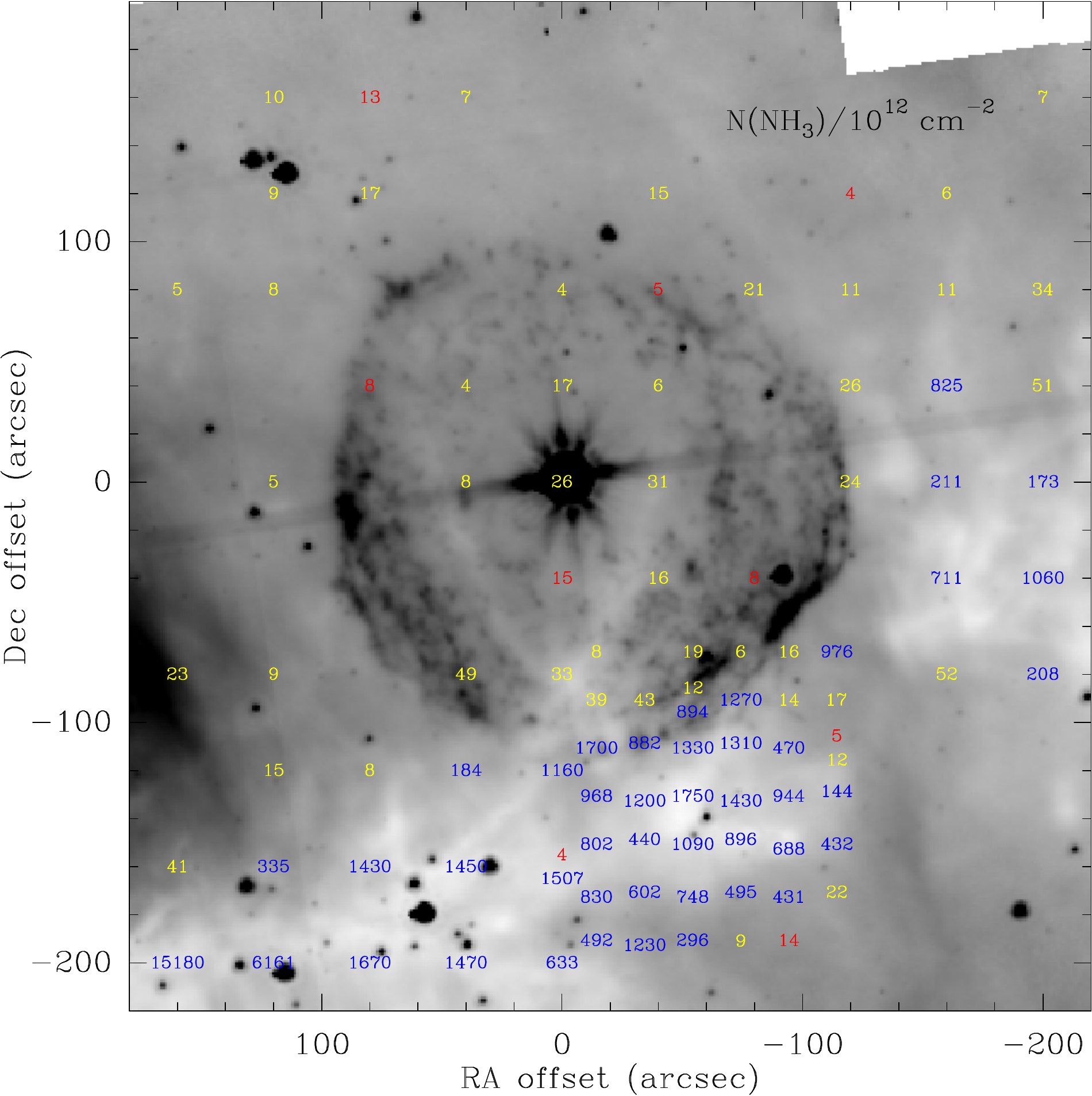}
  \caption{
    Distribution of the ammonia column density, in units of $10^{12}$~\cmt. 
    Colour are the same as Fig.~6. Details about the computation of $N($\nh$)$
    are provided in the text. It is remarkable the large difference between the 
    IRDC and the other positions (see text).
  }
\end{figure}

The general picture of the region consists, therefore, of a very cold and 
massive IRDC, spotted by tenuous areas, some of them depicting moderately high 
temperatures. This strong difference in the temperatures among the IRDC and the 
other positions is even larger if we consider the kinetic temperatures (\tk), 
because the ratio between \tk\ and \trot\ increases with \trot\ 
\citep{dan88,taf04}. At the IRDC, \tk\ is in the range $8-12$\,K, while the 
positions having yellow labels may have \tk\ between 25 and 47\,K. It is hard 
to constrain \tk\ in the hot spots, although large values, 100\,K or even more, 
are possible.

\subsection{Opacities in the IRDC}

Fig.~7 depicts the opacities of the main group ($\tau_m$ in the appendix), 
derived when the hf fitting is possible. The positions correspond to the IRDC, 
and are labelled in blue, as in Fig.~6. Positions having two opacities 
correspond to those cases with two velocity components. As expected in the case 
of dark clouds, the larger opacities are found toward the inner part of the 
IRDC. A notable exception is found in the SW region, where the opacity remains 
high in the northern edge of the region, close to the border of the ring 
nebula. 

\subsection{Column densities and abundances}

Fig.~8 depicts the distribution of the column density, $N($\nh$)$, in the 
observed field. The colour code is the same as in Fig.~6. The estimates of 
$N($\nh$)$ in those hot spots (red labels) where only lower limits to \trot\ 
are determined, was done by assuming \trot~=~50~K; all lower limits to \trot\ 
are higher than 50~K and, therefore, the values of $N($\nh$)$ are lower limits 
in all these cases. At the positions where we determined upper limits (yellow 
labels), we assumed \trot~=~10~K. In those cases having two velocity 
components, the sum of both $N($\nh$)$ are depicted.

After looking at the Fig.~8, the differences between the IRDC and the rest of 
the field are outstanding. $N($\nh$)$ varies from several $10^{14}$ to some 
$10^{16}$ \cmt\ at the IRDC positions, and is in the range from $10^{12}$ to 
$10^{13}$ \cmt\ elsewhere.

This ample range of the $N($\nh$)$ values also affects the estimates of 
abundances with respect to H$_2$, $X($\nh$)$. Fig.~9 shows the $X($\nh$)$ 
distribution in the field, with the same colour codes as previous figures. 
To compute the \nh\ abundance, we had to estimate the H$_2$ column density 
($N($H$_2$) at all positions. This is inferred from the 1.2~mm continuum, 
considered a dust tracer. Assumed dust temperatures were 10~K at the IRDC 
positions and 50~K elsewhere (see the appendix for details about the criteria 
used. 

The lowest values of the abundances, in the range from several $10^{-11}$ to 
some $10^{-9}$, are spread all along the observed field except in the IRDC, 
where the abundance grows up to some $10^{-8}$. 
Interestingly, the abundance in the SW region remains high at its northern and 
south-western borders. This is not the tendency expected in a quiescent clump 
of a dark cloud like, for instance, the other clumps of the IRDC.

\begin{figure}
  \includegraphics[width=\columnwidth]{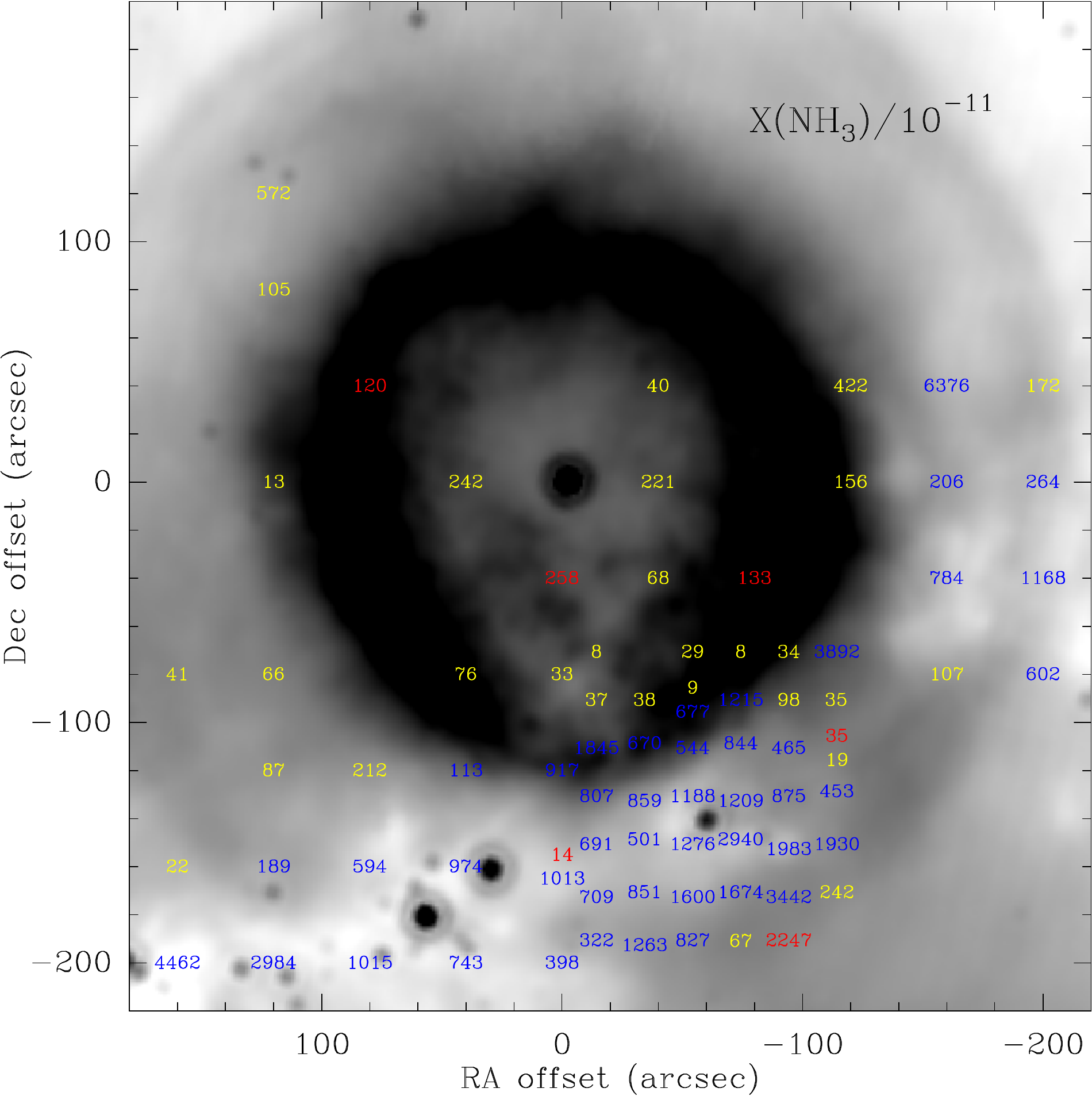}
  \caption{
    Distribution of the ammonia abundances, in units of $10^{-11}$ overlaid to  
    an image of Spitzer/MIPS at 24~$\mu$m to highlight the second infrared 
    ring. See details of the computation of H$_2$ column densities in the text. 
    Notable differences between the IRDC and the rest of the field are again 
    noted; furthermore, high values appear in the northern and south edges of 
    the SW region, coinciding with the position of the first and second 
    infrared rings, respectively (see text).
  }
\end{figure}

\subsection{The \nh\ peak}

In Fig.~10, we show the Boltzmann diagram of the \nh\ peak after fitting the 
(3,3) spectrum to three Gaussian velocity components. The velocity ranges and 
the velocity-integrated line intensities of the fitting are shown in 
Table~\ref{diagrot}. We named the components as central, blue and red according 
to their velocity ranges. 

\begin{table}
\caption{Boltzmann diagram results in the \nh\ peak}
\centering
\begin{tabular}{lrrr}
\hline\hline\noalign{\smallskip}
\multicolumn{1}{c}{Component} & \multicolumn{1}{c}{velocity range} & \multicolumn{1}{c}{\trot} & \multicolumn{1}{c}{$N($\nh$)$} \\
& \multicolumn{1}{c}{\kms} & \multicolumn{1}{c}{K} & \multicolumn{1}{c}{cm$^{-2}$} \\
\hline
\noalign{\smallskip}
Central & $(+0.3, +1.9)$ & $11\pm2$ & $1.7\pm0.2\ 10^{15}$ \\
Blue    & $(-1.7, +0.3)$ & $>40$    & $<1.5\ 10^{12}$ \\
Red     & $(+1.9, +2.8)$ & $>28$    & $<1.5\ 10^{12}$ \\
\hline
\end{tabular}
\label{diagrot}
\end{table}

As said above, the (1,1) and (2,2) lines have not been detected in the red or 
blue velocity components; in these cases, upper detection limits (3-sigma) have 
been computed in the corresponding velocity ranges. 

The results of the Boltzmann diagram are shown in the last two columns of Table 
\ref{diagrot}. \trot\ and $N($\nh$)$ can be determined only for the central 
component. For the others, we only derived lower limits of \trot. As we also 
considered the column densities up to the first four metastable levels (see 
Appendix), the resultant values of $N($\nh$)$ are therefore upper limits.

Most of the gas (of the order of $10^{15}$\,cm$^{-2}$) arises in the central 
component and has a low value of \trot\ (11\,K); these values of \trot\ and 
$N($\nh$)$ are those typically measured in the IRDC.

But strikingly, there is much low amount of gas (three orders of magnitude 
lower) at higher temperatures, above 28\,K and 40\,K. This finding confirms the 
scenario depicted  by the (1,1) and (2,2) lines in the whole field: most of the 
dense gas is in cold form and associated with the IRDC, which in turn is 
embedded in a field dotted by low amounts of warm/hot gas.

\begin{figure}
  \centering
  \includegraphics[width=0.9\columnwidth]{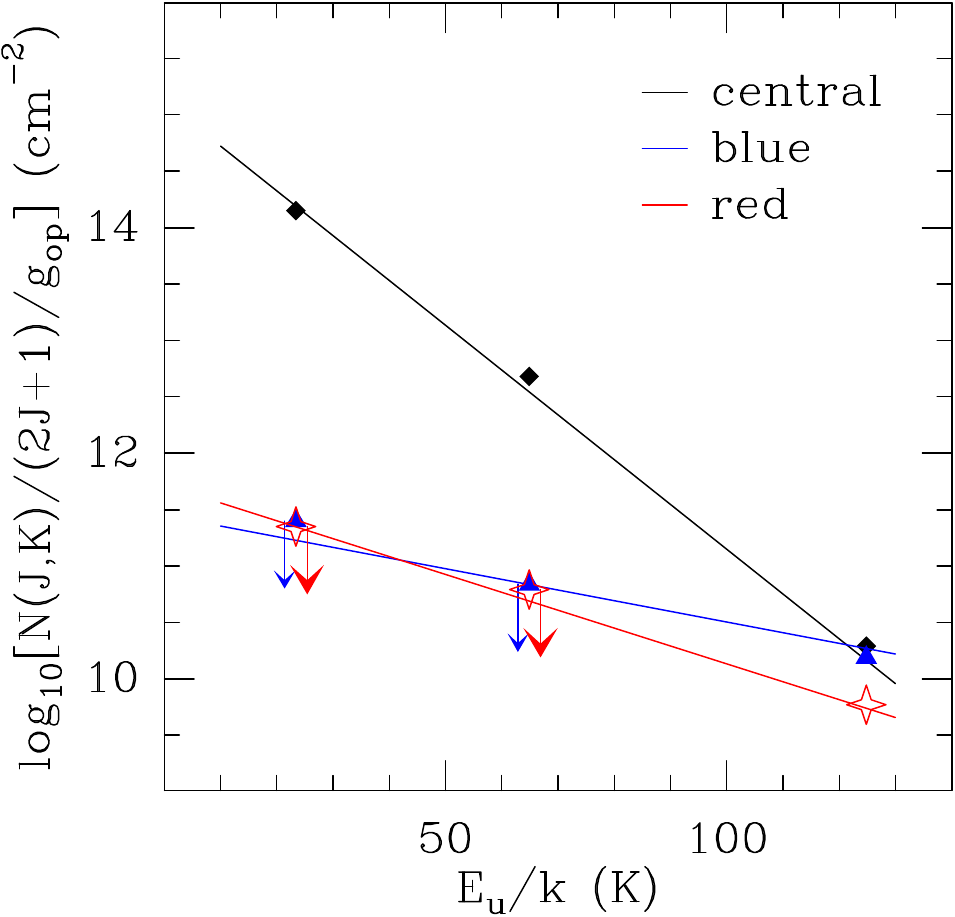}
  \caption{
    Rotational diagram corresponding to the \nh\ peak. The column densities 
    have been computed for the three velocity components depicted in Fig.~5, 
    also indicated in Table 1. The best-fit lines are sketched. The rotational 
    temperature is only determined for the central component, while for the 
    others (blue and red) only lower limits can be derived.
  }
\end{figure}

\subsection{Anomalous \nh(1,1) spectra in the SW region}

We found anomalous satellite ratios in the \nh (1,1) spectra at four positions 
of the SW region. Figure 11 shows the spectra, together with their locations. 
The positions affected by this anomalous emission lie at the outer edges of the 
SW region, two of them close to the CO peak.

The blue outer satellite hyperfine components are stronger than those of the 
red side. This has been proposed to be a signpost of contracting motions, both 
theoretically \citep{par01} and observationally \citep[\eg][]{fon12}.

Alternatively, the anomaly could be due to non-LTE emission due to hyperfine 
selective photon trapping, as this affects only the outer satellites. However, 
this effect would produce the opposite result; the red satellite would become 
stronger than the blue one \citep{stu85}.

\begin{figure*}
  \sidecaption
  \includegraphics[width=12cm]{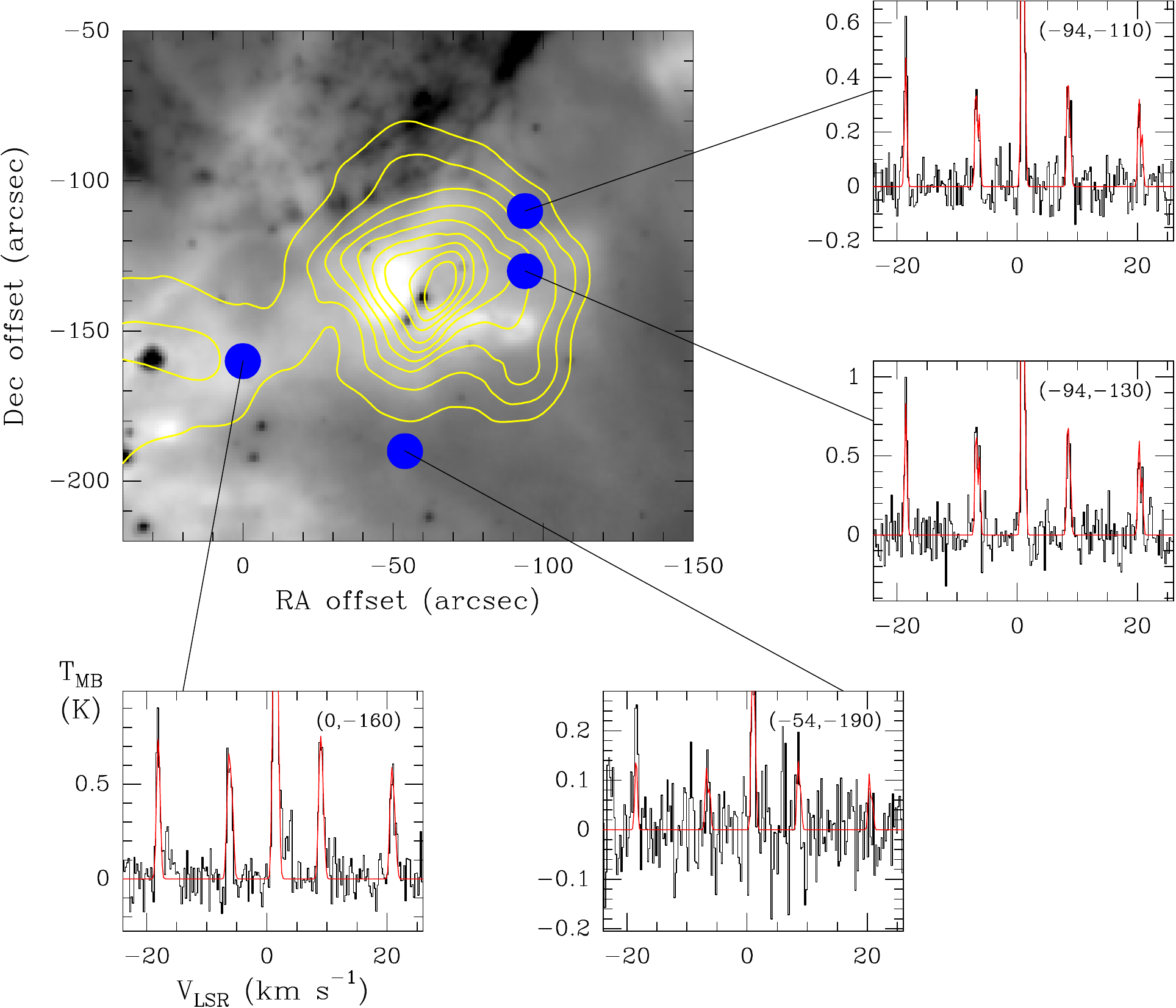}
  \caption{
Anomalous spectra in the SW region. The grey scale displays the Spitzer/IRAC 
8\mum\ emission, and the yellow contours the \nh\ (1,1) integrated emission. 
Blue dots mark the positions of the anomalous spectra. These spectra are shown 
in the smaller boxes outside the map. Offsets are indicated on the top right 
corner of each spectrum. The observed spectra are in black and the fitting in 
red. The $T_\mathrm{MB}$ scale has been chosen to improve the visibility of 
satellite lines.
  }
\end{figure*}

\section{Discussion}
\subsection{Critical revision of the distances}

The most recent distance estimate to some star-forming regions of Cygnus X is 
1.4\,kpc, based on spectroscopic parallaxes of their maser-emitting regions
\citep{ryg12}. This distance is only slightly smaller than previous estimates 
using the same technique \citep{tor91}, and consistent with other works based 
on stellar classification \citep[\eg][]{han03}. On the other hand, several 
studies locate DR15 at a lower distance, about 700-800\,pc 
\citep{wen91,uya01,red03}.

However, the distance to the IRDC is still a matter of debate, with distances 
ranging from 800 to 1700 pc \citep{red03,kra10}. A close inspection of the 
infrared images and submillimeter continuum emission of a large field of view 
($20'\times20'$) including \g\ reveals that the IRDC is interrupted in two 
positions: the SW region \citep[\eg][]{kra10,jim10,uma11}, and the position of 
DR15 \citep[see Fig.~2 of ][]{red03}. At first sight, the IRDC is farther away 
than both DR15 and the LBV nebula. However, the presence of a dark filamentary 
structure extending to the north of the IRDC and crossing the LBV nebula up to 
the position of the LBV star suggests that part of the IRDC is also in front of 
the LBV nebula (Figs.~6, 7, 8, and 11). Therefore, provided that part of the 
IRDC seems to be in front of the LBV nebula, and part of it seems to be behind, 
it seems reasonable to consider that the IRDC and the LBV nebula lie at similar 
distances, of 1.4\,kpc. 

In addition, the overall \nh\ (1,1) and 1.2~mm emission shown in Fig.~1 reveal 
a discontinuity separating the SW region and the westernmost clump of the IRDC, 
which is coincident with a bright 8~$\mu$m structure extending from the 
infrared ring farther to the south-west (Figs.~6, 7, 8, and 11). This has been 
proposed as an interaction zone of the LBV star with the surrounding medium 
\citep{uma11} and favours the association of the LBV star with the IRDC.

The finding of hot spots with low column density in the field of view toward 
\g\ leads us to associate the most tenuous ammonia to the Cygnus~X 
star-forming region, or even to \g\ itself. In this scenario, the warm/hot 
spots may represent the relics of the clumps of the ancient molecular cloud, 
which are being photo-evaporated by the UV field from \g, perhaps supported by 
other neighbouring massive stars. This idea is reinforced by the low abundance 
of \nh\ measured near \g, compatible with the idea of \nh\ being 
photo-dissociated by the strong UV radiation present in the region. The high 
temperatures found in the nebula are consistent with the high dust temperature 
already found in the same region \citep{wat96,jim10}. 

In summary, we are providing here sufficient arguments which favour the 
possibility of some co-existence of the IRDC and \g\ in close volumes. 
Therefore, the interplay between these two objects should be explored in more 
detail.

\subsection{Hints of a possible interaction in the SW region}

As said above, the fraction of the IRDC which we mapped is roughly divided in 
four regions, more or less aligned from the south-east corner of the map to its 
western border. In some aspects, the SW region (one of these four regions) 
presents a number of peculiarities as compared to the other three regions.

The opacities measured in the SW region do not clearly decrease at the borders, 
which is not the expected behaviour in a typical cold dark cloud. In 
particular, the opacity in the northern edge --roughly a strip from 
$(0", -120")$ to $(-80", -100")$-- remains as high as in the centre. The 
abundance is also high in some points of the same strip, up to one order of 
magnitude larger compared to the \nh\ peak (located at the centre of the SW 
region). The southern border --a strip from $(-30", -200")$ to 
$(-120", -150")$-- also depicts some points of high relative abundance. The 
anomalous (1,1) satellite emission (Sect.~4.6) in the external parts of the SW 
region is also another peculiarity.

The large differences in some parameters of the SW region, in comparison to the 
rest of the IRDC, should be due to different physical conditions. An increase 
of the abundances towards the borders is not the expected tendency in the parts 
of a cold cloud most exposed to external UV radiation. The abundance of \nh\ in 
a harsh environment may be enhanced by sputtering in the presence of a C-type 
shock \citep{flo94, dra95}, which releases volatile molecules from the surface 
of the dust grains. This is a reasonable hypothesis taking into account the 
presence of a low-velocity shock (about 14\kms) at the north-western side of 
the SW region, close to the CO peak \citep{riz08}. It is also worth noting that 
the northern edge of the SW region coincides with a brightening and 
morphological disturbance of the 6-cm continuum emission, which is claimed as 
an evidence of some interaction with the surrounding ISM \citep{uma11}. 
Therefore, these ammonia observations may have unveiled the molecular 
counterpart of such disturbance, and also a new molecular component associated 
with the shocks discovered in CO \citep{riz08}.

The positions of abundance enhancement towards the southern border of the SW 
region agree with those of the external infrared shell, better seen in 24\mum\ 
\citep[][and Fig.~9 of this work]{jim10}.

The scenario of the LBV star interacting with the IRDC is also consistent with 
the (3,3) spectrum towards \nh\ peak, which reveals two velocity components 
tracing gas at temperatures  $\gtrsim30$\,K, clearly larger than the typical 
temperatures measured in the IRDC, of $\sim10$\,K.

Therefore, a number of observational findings points to a possible interaction 
of the nebula and the nested molecular shells with the SW region. A definitive 
evidence will come from observations with high angular resolution.

\subsection{Triggered star formation in the SW region}

If the LBV star is interacting with the SW region this could affect the star 
formation which is currently ongoing in the area \citep{vin08}. The dynamical 
time scale of the ejection events which produced the CO and infrared nested 
shells is $10{^4}$--$10{^5}$ yr \citep{riz08,uma11}. This is comparable to the 
age of the brightest young stellar object embedded in the SW region, assuming 
it is in the Class 0 phase \citep{gre94,ken95,eno09,eva09}. Thus, while it 
seems unlikely that the ejection events of the LBV could have triggered the 
collapse of the young stellar objects embedded in the SW region, an interesting 
possibility would be to explore whether they are affecting the formation and 
evolution of the young stellar objects already formed.

After some $10^4-10^5$\,yr of evolution, \g\ will become a Wolf-Rayet star and 
the \nh\ ring will eventually disappear. The stellar wind regime will change 
at this stage, turning it out to higher velocity and lower wind density 
\citep{lan94,mae94,gar95}. This will produce new shocks in the surrounding gas 
and dust, which may release volatile molecules from the icy grain mantles 
\citep{flo94} and consequently increase the ammonia abundance again. This line 
of reasoning is supported by the significantly high \nh\ abundance seen to the 
north and south-western borders of the SW region, a trend also found in the 
molecular cloud around the WR nebula NGC\,2359 \citep{riz01a}, and the Galactic 
Center \citep{flo95,mar99}. The young stellar objects formed in the SW region 
will probably suffer photo-evaporation and heating of their envelopes.

\section{Conclusions}

In order to study the molecular gas in the infrared ring nebula of the LBV star 
\g\ and its nearby IRDC, we used the Effelsberg telescope to map the \nh\ (1,1) 
and \nh(2,2) emission in the \g\ field. We additionally observed \nh(3,3) in 
particular positions of the strongest \nh(1,1) emission. Our main findings are 
summarized as follows:

\begin{enumerate}

\item 
While the strongest emission of \nh\ (1,1) and (2,2) matches the infrared-dark 
structure of the IRDC, mainly detected at velocities $>1$\kms, the emission in 
the velocity range from $-1.2$\kms\ to $+0.5$\kms\ reveals a coherent structure 
closely following the infrared ring surrounding the LBV star, also partially 
seen in the (2,2) transition. This is the first \nh\ structure associated with 
an evolved massive star known so far.

\item 
We derived the rotational temperature and \nh\ column density of the dense gas 
in the entire field of view, and found that the rotational temperature is 
uniform and in the range 7--11\,K in the IRDC, while it is $>30$\,K in 
particular positions or hot spots of the infrared ring nebula and a couple of 
points within $30''$ of the LBV star. The \nh\ column density in the IRDC is in 
the range 10$^{14}$--10$^{15}$\,cm$^{-2}$, while it is about two orders of 
magnitude lower near the LBV and the infrared ring. Using the 1.2\,mm continuum 
emission in the field, we inferred \nh\ abundances of about 10$^{-8}$ in the 
IRDC, and 10$^{-10}$--10$^{-9}$ near the LBV star. The warm temperatures and 
low abundances of \nh\ near the LBV star and its infrared-ring suggest that the 
gas is being heated and photo-dissociated by the intense UV-field of the LBV 
star.

\item 
An outstanding region is recognized beyond the ring nebula towards its 
south-western part, which is also part of the IRDC (the SW region). The \nh\ 
(3,3) emission towards the centre of the SW region reveals three velocity 
components. One of them is associated with the IRDC and has a rotational 
temperature of 11~K, while the other two are associated with small dense cores 
with temperatures $>30$\,K. In the northern edge of the SW region, the opacity 
of the (1,1) line keeps a value as high as in the clump centre. The \nh\ 
abundance tends to be higher at the clump edges than in the clump centre. All 
these features, together with the results of previous works reporting a shock 
within this region and hints of interaction with the LBV star, strongly 
suggests that a mass-loss ejection event of the LBV star is interacting with 
the SW region and releasing \nh\ molecules back to the gas phase, thus 
increasing its abundance at the outer edges of the clump.

\end{enumerate}

\g\ is providing increasing evidences of being an excellent laboratory to 
study the history of a high-mass star evolution. Being LBVs the precursors of 
core-collapse supernovae (through a Wolf-Rayet stage), the knowledge of the 
different actors present in this field may also help to learn about this topic. 
The ammonia ring-like structure associated to \g\ and the SW region deserve 
follow up observations with high angular resolution. Furthermore, the 
observation of other molecular tracers would be of particular interest.

\begin{acknowledgements}

We are deeply grateful to Gemma Busquet and \'Alvaro S\'anchez-Monge for 
observing the final part of the map. We also acknowledge the kind and 
professional support of the Effelsberg staff during the observations. The 
constructive comments from the anonymous referee and the critical reading of 
the editor have significantly improved the paper. JRR acknowledges support 
from MICINN (Spain) grants CSD2009-00038, AYA2009-07304, and AYA2012-32032. AP 
is supported by a JAE-Doc CSIC fellowship co-funded with the European Social 
Fund under the program `Junta para la Ampliaci\'on de Estudios', by the MICINN 
grant AYA2011-30228-C03-02 (co-funded with FEDER funds), and by the AGAUR grant 
2009SGR1172 (Catalonia). FJ-E acknowledges support from MICINN grant 
AYA2011-24052 and the CoSADIE Coordination Action (FP7, Call INFRA-2012-3.3 
Research Infrastructures, project 312559).

\end{acknowledgements}


\begin{thebibliography}{}

\bibitem[Busquet et al.(2009)]{bus09} Busquet, G., Palau, A., Estalella, 
R., et al.\ 2009, \aap, 506, 1183 

\bibitem[Dame \& Thaddeus(1985)]{dam85} Dame, T.~M., \& Thaddeus, P.\ 1985, 
\apj, 297, 751 

\bibitem[Danby et al.(1988)]{dan88} Danby, G., Flower, D.~R., Valiron, P., 
Schilke, P., \& Walmsley, C.~M.\ 1988, \mnras, 235, 229 

\bibitem[Draine(1995)]{dra95} Draine, B.~T.\ 1995, \apss, 233, 111 

\bibitem[Enoch et al.(2009)]{eno09} Enoch, M.~L., Evans, N.~J., 
II, Sargent, A.~I., \& Glenn, J.\ 2009, \apj, 692, 973

\bibitem[Evans et al.(2009)]{eva09} Evans, N.~J., II, Dunham, 
M.~M., J{\o}rgensen, J.~K., et al.\ 2009, \apjs, 181, 321

\bibitem[Flower \& Pineau des For\^ets(1994)]{flo94} Flower, D.~R., \& 
Pineau des For\^ets, G.\ 1994, \mnras, 268, 724

\bibitem[Flower et al.(1995)]{flo95} Flower, D.~R., Pineau des For\^ets, G., 
\& Walmsley, C.~M.\ 1995, \aap, 294, 815 

\bibitem[Fontani et al.(2012)]{fon12}
Fontani, F., Palau, A., Busquet, G., et al.\ 2012, \mnras, 423, 1691

\bibitem[Fuente et al.(1990)]{fue90} Fuente, A., Martin-Pintado, J., 
Bachiller, R., \& Cernicharo, J.\ 1990, \aap, 237, 471 

\bibitem[Gal-Yam et al.(2007)]{gal07} Gal-Yam, A., Leonard, 
D.~C., Fox, D.~B., et al.\ 2007, \apj, 656, 372 

\bibitem[Garcia-Segura \& Mac Low(1995)]{gar95} Garcia-Segura, G., \& Mac Low, 
M.-M.\ 1995, \apj, 455, 145 

\bibitem[Goldsmith \& Langer(1999)]{gol99} Goldsmith, P.~F., \& Langer, W.~D.\ 
1999, \apj, 517, 209 

\bibitem[Gottschalk et al.(2012)]{got12} Gottschalk, M., Kothes, R., 
Matthews, H.~E., Landecker, T.~L., \& Dent, W.~R.~F.\ 2012, \aap, 541, A79 

\bibitem[Greene et al.(1994)]{gre94} Greene, T.~P., Wilking, B.~A., 
Andre, P., Young, E.~T., \& Lada, C.~J.\ 1994, \apj, 434, 614

\bibitem[Guesten et al.(1985)]{gue85} G\"usten, R., Walmsley, C.~M., 
Ungerechts, H., \& Churchwell, E.\ 1985, \aap, 142, 381 

\bibitem[Hanson(2003)]{han03}
Hanson, M.~M.\ 2003, \apj, 597, 957

\bibitem[Higgs et al.(1994)]{hig94} Higgs, L.~A., Wendker, H.~J., \& Landecker, 
T.~L.\ 1994, \aap, 291, 295 

\bibitem[Ho \& Townes(1983)]{ho83} Ho, P.~T.~P., \& Townes, C.~H.\ 1983, 
\araa, 21, 239 

\bibitem[Jijina et al.(1999)]{jij99} Jijina, J., Myers, 
P.~C., \& Adams, F.~C.\ 1999, \apjs, 125, 161 

\bibitem[Jim\'enez-Esteban et al.(2010)]{jim10} Jim\'enez-Esteban, 
F.~M., Rizzo, J.~R., \& Palau, A.\ 2010, \apj, 713, 429 

\bibitem[Kenyon \& Hartmann(1995)]{ken95}
Kenyon, S.~J., \& Hartmann, L.\ 1995, \apjs, 101, 117

\bibitem[Kraemer et al.(2010)]{kra10}
Kraemer, K.~E., Hora, J.~L., Egan, M.~P., et al.\ 2010, \aj, 139, 2319

\bibitem[Langer et al.(1994)]{lan94} Langer, N., Hamann, W.-R., Lennon, M., 
et al.\ 1994, \aap, 290, 819 

\bibitem[Loinard et al.(2012)]{loi12} Loinard, L., Menten, K.~M., G{\"u}sten, 
R., Zapata, L.~A., \& Rodr{\'{\i}}guez, L.~F.\ 2012, \apjl, 749, L4 

\bibitem[Longmore et al.(2007)]{lon07}
Longmore, S.~N., Burton, M.~G., Barnes, P.~J., et al.\ 2007, \mnras, 379, 535

\bibitem[Maeder \& Meynet(1994)]{mae94} Maeder, A., \& Meynet, G.\ 1994, 
\aap, 287, 803 

\bibitem[Maret et al.(2009)]{mar09} 
Maret, S., Faure, A., Scifoni, E., \& Wiesenfeld, L.\ 2009, \mnras, 399, 425 

\bibitem[Mart{\'{\i}}n-Pintado et al.(1999)]{mar99} 
Mart{\'{\i}}n-Pintado, J., Gaume, R.~A., Rodr{\'{\i}}guez-Fern{\'a}ndez, 
N., de Vicente, P., \& Wilson, T.~L.\ 1999, \apj, 519, 667 

\bibitem[Motte et al.(1998)]{mot98} Motte, F., Andre, P., \& Neri, R.\ 
1998, \aap, 336, 150 

\bibitem[Motte et al.(2007)]{mot07} Motte, F., Bontemps, S., Schilke, P., 
Schneider, N., Menten, K.~M., Brogui\`ere, D. 2007, A\&A, 476, 1243

\bibitem[Ossenkopf \& Henning(1994)]{oss94} Ossenkopf, V., \& Henning, T.\ 
1994, \aap, 291, 943 

\bibitem[Ott et al.(1994)]{ott94} Ott, M., Witzel, A., Quirrenbach, A., 
et al.\ 1994, \aap, 284, 331 

\bibitem[Palau et al.(2007)]{pal07} Palau, A., Estalella, R., Girart, J.~M., 
et al.\ 2007, \aap, 465, 219 

\bibitem[Park(2001)]{par01}
Park, Y.-S.\ 2001, \aap, 376, 348

\bibitem[Pauls et al.(1983)]{pau83} Pauls, A., Wilson, T.~L., Bieging, 
J.~H., \& Martin, R.~N.\ 1983, \aap, 124, 23 

\bibitem[Poglitsch et al.(2010)]{pog10} Poglitsch, A., Waelkens, C., Geis, 
N., et al.\ 2010, \aap, 518, L2 

\bibitem[Redman et al.(2003)]{red03} Redman, R.~O., Feldman, 
P.~A., Wyrowski, F., et al.\ 2003, \apj, 586, 1127

\bibitem[Rizzo et al.(2003a)]{riz03a} Rizzo, J.~R., Mart{\'{\i}}n-Pintado, 
J., \& Desmurs, J.-F.\ 2003a, \aap, 411, 465 

\bibitem[Rizzo et al.(2003b)]{riz03b} Rizzo, J.~R., Mart{\'{\i}}n-Pintado, 
J., \& Desmurs, J.-F.\ 2003b, IAU Symp., 212, 742 

\bibitem[Rizzo et al.(2008)]{riz08} Rizzo, J.~R., 
Jim\'enez-Esteban, F.~M., \& Ortiz, E.\ 2008, \apj, 681, 355 

\bibitem[Rizzo et al.(2001a)]{riz01a} Rizzo, J.~R., 
Mart{\'{\i}}n-Pintado, J., \& Henkel, C.\ 2001a, \apjl, 553, L181 

\bibitem[Rizzo et al.(2001b)]{riz01b} Rizzo, J.~R., 
Mart{\'{\i}}n-Pintado, J., \& Mangum, J.~G.\ 2001b, \aap, 366, 146

\bibitem[Rygl et al.(2012)]{ryg12} Rygl, K.~L.~J., Brunthaler, A., Sanna, A., 
et al.\ 2012, \aap, 539, A79 

\bibitem[Schneider et al.(2006)]{sch06} Schneider, N., Bontemps, S., 
Simon, R., et al.\ 2006, \aap, 458, 855 

\bibitem[Sep{\'u}lveda et al.(2011)]{sep11} Sep{\'u}lveda, I., Anglada, G., 
Estalella, R., et al.\ 2011, \aap, 527, A41 

\bibitem[Smith(2007)]{smi07} Smith, N.\ 2007, \aj, 133, 1034 

\bibitem[Smith et al.(2006)]{smi06} Smith, N., Brooks, K.~J., 
Koribalski, B.~S., \& Bally, J.\ 2006, \apjl, 645, L41 

\bibitem[Smith et al.(1998)]{smi98} Smith, L.~J., Nota, A., 
Pasquali, A., et al.\ 1998, \apj, 503, 278 

\bibitem[Stutzki \& Winnewisser(1985)]{stu85}
Stutzki, J., \& Winnewisser, G.\ 1985, \aap, 144, 13

\bibitem[Tafalla \& Bachiller(1995)]{taf95} Tafalla, M., 
\& Bachiller, R.\ 1995, \apjl, 443, L37 

\bibitem[Tafalla et al.(2004)]{taf04} 
Tafalla, M., Myers, P.~C., Caselli, P., \& Walmsley, 
C.~M.\ 2004, \aap, 416, 191 

\bibitem[Torres-Dodgen et al.(1991)]{tor91}
Torres-Dodgen, A.~V., Carroll, M., \& Tapia, M.\ 1991, \mnras, 249, 1

\bibitem[Townes \& Schawlow(1975)]{tow75} Townes, C.~H., \& Schawlow, A.~L.\ 
1975, Microwave spectroscopy., New York: Dover Publications

\bibitem[Umana et al.(2011)]{uma11} Umana, G., Buemi, C.~S., 
Trigilio, C., et al.\ 2011, \apjl, 739, L11

\bibitem[Ungerechts et al.(1986)]{ung86} Ungerechts, H., Winnewisser, G., \& 
Walmsley, C.~M.\ 1986, \aap, 157, 207 

\bibitem[Uyan{\i}ker et al.(2001)]{uya01}
Uyan{\i}ker, B., F{\"u}rst, E., Reich, W., Aschenbach, B., \& 
Wielebinski, R.\ 2001, \aap, 371, 675

\bibitem[Vink et al.(2008)]{vin08} Vink, J.~S., Drew, J.~E., Steeghs, D., 
et al.\ 2008, \mnras, 387, 308 

\bibitem[Waters et al.(1996)]{wat96} Waters, L.~B.~F.~M., Izumiura, H., Zaal, 
P.~A., et al.\ 1996, \aap, 313, 866 

\bibitem[Wendker et al.(1991)]{wen91}
Wendker, H.~J., Higgs, L.~A., \& Landecker, T.~L.\ 1991, \aap, 241, 551

\bibitem[Wilson et al.(2009)]{wil09} Wilson, T.~L., Rohlfs, K., 
H{\"u}ttemeister, S.\ 2009, Tools of Radio Astronomy, Springer-Verlag, Berlin

\bibitem[Zhang et al.(2002)]{zha02} Zhang, Q., Hunter, T.~R., Sridharan, T.~K., 
\& Ho, P.~T.~P.\ 2002, \apj, 566, 982 

\end{thebibliography}


\clearpage
\begin{appendix}

\section{Channel maps}

The proper channel maps of the two ammonia lines observed are presented in 
Figs.~A.1 and A.2. The collection of spectra are also available at CDS.

\begin{figure*}
  \centering
  \includegraphics[width=0.95\textwidth]{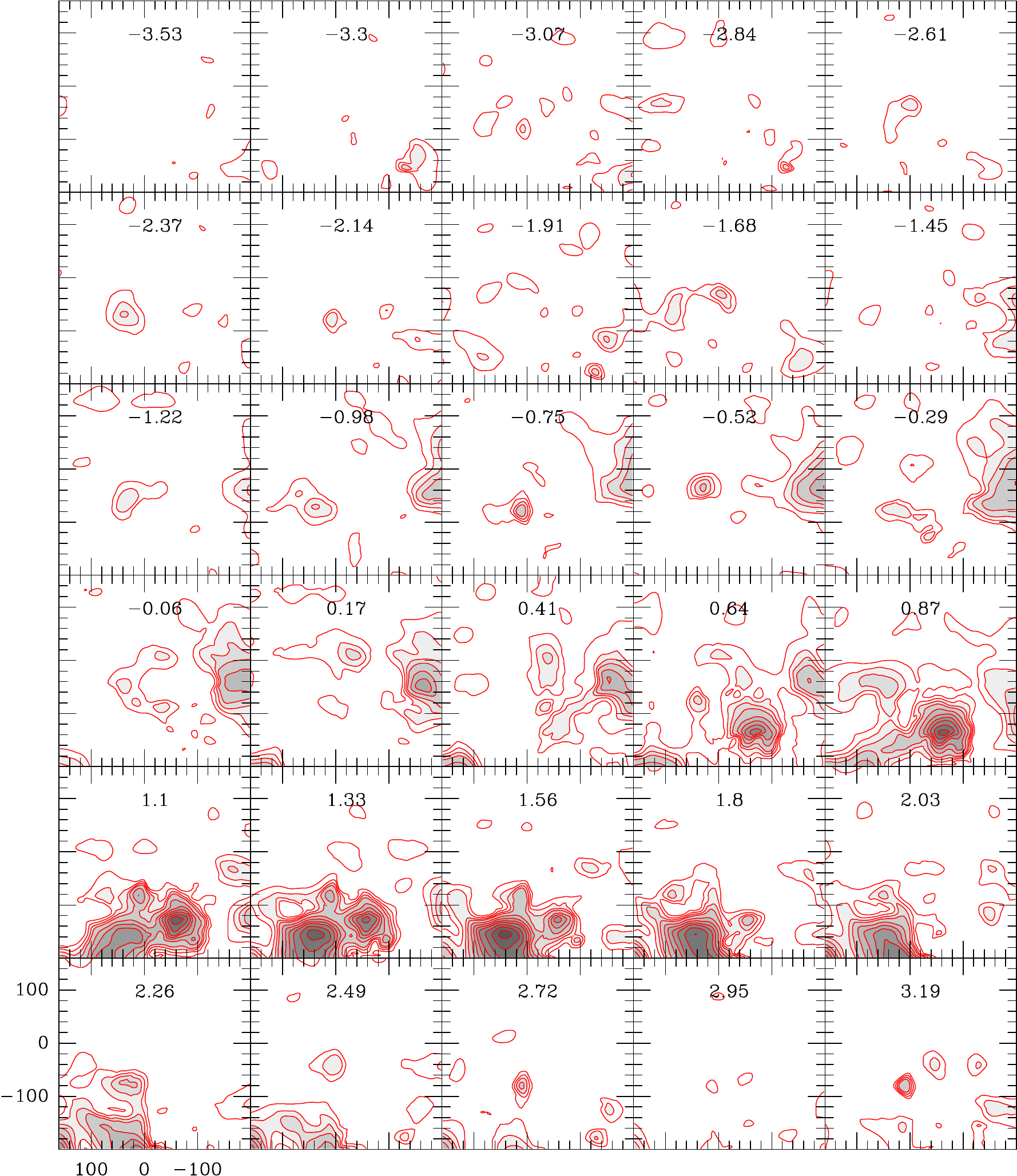}
  \caption{
    Channel maps corresponding to the \nh\ (1,1) emission. Only the main 
    hyperfine component is depicted. LSR velocities are indicated at the top 
    center of each map. Contours are 0.15, 0.30, 0.45, 0.60, 0.90, 1.20, 1.50, 
    1.80, 2.25, 2.70, 3.15, 3.60, 4.20, and 4.8~K.
  }
\end{figure*}

\begin{figure*}
  \centering
  \includegraphics[width=0.95\textwidth]{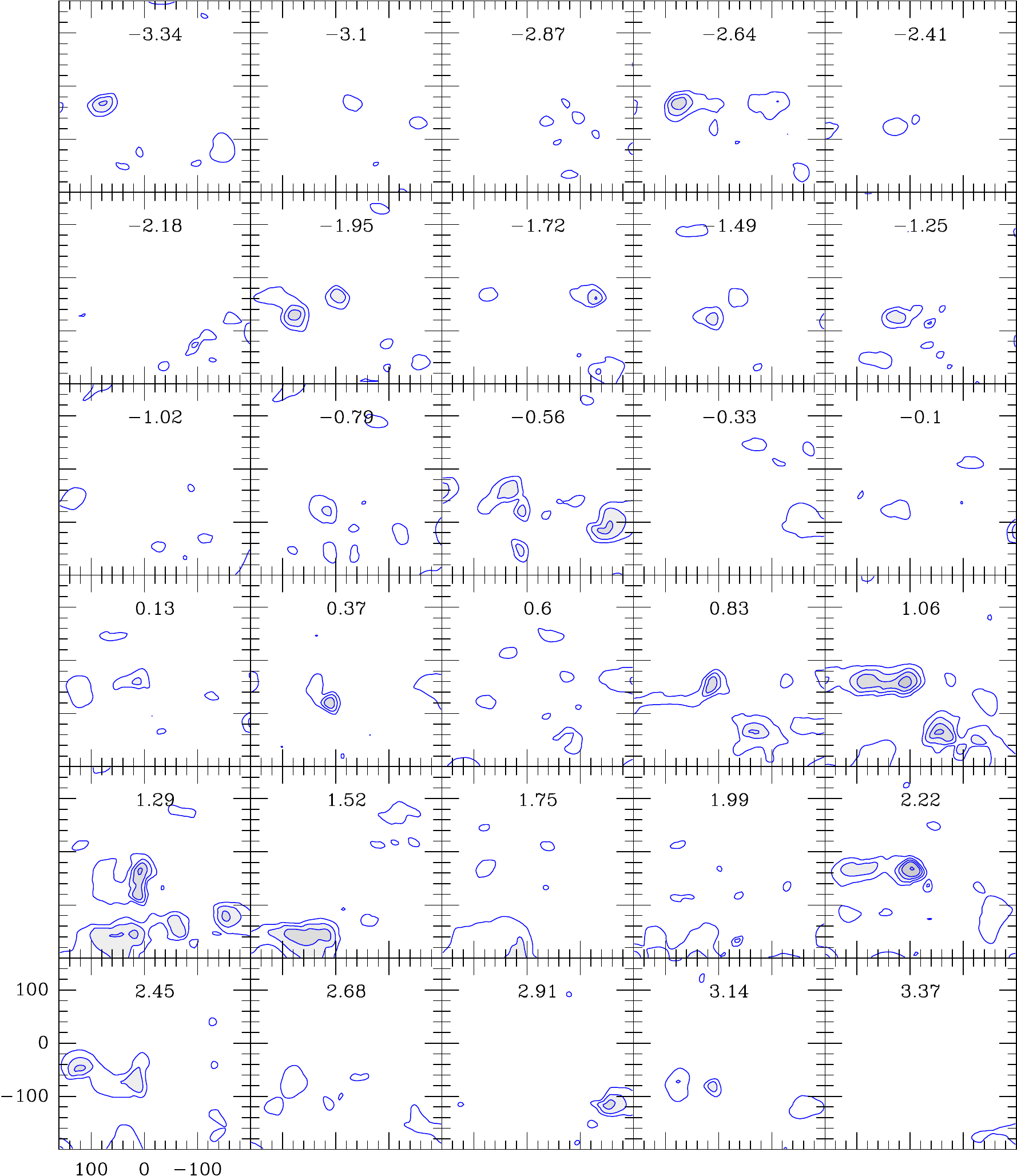}
  \caption{
    Channel maps corresponding to the \nh\ (2,2) emission. Only the main 
    hyperfine component is depicted. LSR velocities are indicated at the top 
    center of each map. Contours are 0.18, 0.36, 0.54, 0.72, and 1.08~K.
  }
\end{figure*}

\clearpage

\section{Formulation}

In this appendix we summarize the formula used for the determination of some 
physical properties (opacities, rotational temperatures, column densities, and 
abundances) from the observed \nh\ lines. This corresponds to the standard 
interpretation already discussed and presented by several authors, such as 
\citet{ung86} and \citet{bus09}.

\subsection{Hyperfine and Gaussian fitting}

The \nh\ lines are splitted by the quadrupole hyperfine interaction 
\citep{ho83}. From the relative intensities of the main hf group and the four 
satellite ones, it is possible to directly derive the optical depth of the main 
hf group \citep{pau83}. When the satellite (1,1) lines are detected, a CLASS 
method was used for this fitting. The output are: 
(1) $A\,\tau_\mathrm{m}=f\,[J_\nu(T_\mathrm{exc})-J_\nu(T_\mathrm{bg})]\,\tau_\mathrm{m}$; 
(2) LSR velocity ($V_\mathrm{LSR}$); 
(3) line width at half maximum ($\Delta v$); and 
(4) $\tau_\mathrm{m}$, the opacity of the main hf group. 
$f$ is the beam filling factor, and $J_\nu(T)$ is Rayleigh-Jeans temperature, 
defined as 
$J_\nu(T)=({\mathrm h}\,\nu / {\mathrm k})~(e^{{\mathrm h}\,\nu / {\mathrm k}\,T}-1)^{-1}$, 
\tbg~=~2.73~K is the background temperature, and h and k are the Planck and 
Boltzmann constants, respectively.

This formulation assumes that all the hf levels are populated according to 
local thermodynamic equilibrium (LTE) conditions, and a Gaussian velocity 
distribution. The level population is characterized by the excitation 
temperature \texc, which can be cleared from the fitting by:

\begin{equation}
T_\mathrm{exc}=\frac{{\mathrm h} \nu / {\mathrm k}}{ln[1+{\mathrm h} \nu / {\mathrm k}\ \ W^{-1}]}\,,
\end{equation}

\noindent 
where $W = J_\nu(T_\mathrm{exc}) = A/f+J_\nu(T_\mathrm{bg})$. 

The column density of a \nh\ $(J,K)$ level, $N(J,K)$, is obtained by:

\begin{equation}
N(J,K)=\frac{1.65\,10^{14}}{\nu}\ \frac{J (J+1)}{K^2}\ \Delta v\ T_\mathrm{exc}\ \tau_\mathrm{tot}\ \mathrm{cm}^{-2}\,,
\end{equation}

\noindent
where $\nu$ is the line frequency in GHz, $\tau_\mathrm{tot}$ is the ``total'' 
opacity of all the hf components, and $\Delta v$ and \texc\ are in \kms\ and K, 
respectively. The above equation results from solving the Transport Equation 
for the \nh\ molecule \citep{wil09}, using a dipole moment $\mu=1.469$~D, and 
assuming \texc\ $>>$ \tbg.

When the hyperfine fitting is possible, we estimate the column density of the 
(1,1) level, N(1,1), using the above equation. Numerically, it is computed by:

\begin{equation}
N(1,1)= 2.785\ 10^{13}\ \Delta v\ T_\mathrm{exc}\ \tau_\mathrm{m}\ \mathrm{cm}^{-2}\,,
\end{equation}

\noindent
where a rough assumption of $\tau_\mathrm{tot}=2\ \tau_\mathrm{m}$ is used.

When the hyperfine fitting is not possible, we have assumed optically thin 
emission. In this case, Eq.~(B.2) for the (1,1) line turns out to:

\begin{equation}
N(1,1)=1.308\ 10^{13}\ \int{T_\mathrm{MB}\ dv}\ \mathrm{cm}\,^{-2}\,.
\end{equation}

In the above equation, all the hf components are included in the integral. If 
the integration extends only to the main group, the coefficient should then be 
multiplied by a factor of 2.

In the case of the (2,2) line, the equivalent expression is:

\begin{equation}
N(2,2)=9.801\ 10^{12}\ \int{T_\mathrm{MB}\ dv}\ \mathrm{cm}\,^{-2}\,.
\end{equation}

Provided $N(1,1)$ and $N(2,2)$, the rotational temperature \trot\ can be 
computed \citep{tow75, ho83} by:

\begin{equation}
T_\mathrm{rot}=\frac{-41.5}{ln\left(\frac{3}{5}\,\frac{N(2,2)}{N(1,1)}\right)}~{\mathrm K}\,,
\end{equation}

\noindent
which results from a two-level system approximation \citep{ho83}.

The partition function is given by

\begin{equation}
\begin{array}{rrl}
Q(T)&\!\!\!=\!\!\!&\sum\limits_{J,K}\ g_{JK}\ e^{-E_{JK}/\mathrm{k}T} \\ 
\\
&\!\!\!\approx\!\!\!& 1+3\,e^{-23.4/T}+5\,e^{-64.9/T}+14\,e^{-124.8/T}\,,
\end{array}
\end{equation}

\noindent
where $g_{JK}=(2J+1)\,g_{op}$ and $E_{JK}$ are the statistical weight and the 
energy of the level $(J,K)$, and $g_{op}$ is the statistical weight for the 
ortho and para species. In the above equation we summed up to the first four 
levels\,\footnote{
The error introduced by this simplification is lower than 
$8\ 10^{-7}$~\%, 1.3~\%, 3~\%, and 17~\% for 
\tk~=~10~K, 20~K, 50~K, and 100~K.
}
, and assumed that (1) only the metastable levels are populated; (2) the levels 
are characterized by the LTE temperature T; and (3) $g_{op}$ equals to 2 and 1 
for the ortho and para species, respectively.

The total column density of all the metastable lines may be computed from a 
single line and by assuming a partition function characterized by 
$T$\,=\,\trot \citep[\eg][]{ung86,bus09}:

\begin{equation}
N(\mathrm{NH}_3)=\left(N(J,K)/g_{JK}\right)\ Q(T_\mathrm{rot})\ e^{E_{JK}/\mathrm{k}T_\mathrm{rot}}.
\end{equation}

In particular, for the (1,1) line the above equation becomes

\begin{equation}
N(\mathrm{NH}_3)=N(1,1)\left[\frac{1}{3}e^{23.4/T_\mathrm{rot}}+1+
\frac{5}{3}e^{-41.5/T_\mathrm{rot}}+\frac{14}{3}e^{-101.2/T_\mathrm{rot}}\right].
\end{equation}

When the (2,2) line is the only one detected, an equivalent formula to (B.9) is:

\begin{equation}
N(\mathrm{NH}_3)=N(2,2)\left[\frac{1}{5}e^{64.9/T_\mathrm{rot}}+
\frac{3}{5}e^{41.5/T_\mathrm{rot}}+1+\frac{14}{5}e^{-59.6/T_\mathrm{rot}}\right].
\end{equation}

\subsection{Abundances}

The \nh\ abundance is computed by

\begin{equation}
X(\mathrm{NH}_3)=N(\mathrm{NH}_3)/N(\mathrm{H}_2)\,,
\end{equation}

\noindent where $N($H$_2)$ is the H$_2$ column density. To estimate it, we used 
the survey of the Cygnus region done by \citet{mot07} in the 1.2\,mm 
continuum emission.

At this wavelength, most of the flux arises from thermal dust emission, which 
is optically thin. By assuming a constant gas-to-dust ratio, the 1.2\,mm flux 
results directly related to the total gas. We therefore used the formulation of 
\citet{mot98}, accordingly adapted to our case.

More precisely, we have used their Eq.~(1'), which compute $N($H$_2)$ as a 
function of the 1.2\,mm flux, the dust temperature ($T_{\mathrm {dust}}$) and 
$\kappa_{\mathrm {1.2mm}}$, the dust opacity per unit mass column density.

The map of \citet{mot07} was convolved to 40\arcsec\ in order to match the 
100m's angular resolution. For $\kappa_{\mathrm {1.2mm}}$ we have adopted a 
value of 0.01~g$^{-1}$\,(cm$^{-2}$)$^{-1}$, which corresponds to the case of 
dust particles covered by thin ice mantles \citep{oss94}, and is also a 
geometrical average between the values usually adopted for pre-stellar dense 
clumps and circumstellar envelopes in young stellar objects of class II 
\citep{mot07}.

Under these assumptions, the resulting formula is 

\begin{equation}
N({\mathrm H}_2)\approx2\,10^{20}~S({\mathrm{mJy}})~\left(\frac{T_{\mathrm
{dust}}}{20\,{\mathrm K}}\right)^{-1}~{\mathrm{cm}}^{-2}\,,
\end{equation}

\noindent
where $S$ is the flux at 1.2\,mm in the convolved map. $T_{\mathrm {dust}}$ is 
assumed as 10\,K in the IRDC positions, and 50\,K elsewhere,
consistent with the two dust population previously reported in the region
\citep{uma11}.

For the IRDC, the assumed value of 10\,K is compatible with the \trot\ derived 
from the hf fitting, because \tk$\approx$\trot\ at low temperatures, and the 
gas is thermalised at the dust temperature. At the other positions 
--particularly in the ring nebula and its interior-- the continuum emission is 
bright at shorter wavelengths, such as 24\mum, indicating a dust temperature 
probably larger. By assuming a conservative value of 50\,K the derived values 
of $N($H$_2)$ have to be considered as upper limits and thus $X($\nh$)$ as 
lower limits.

\subsection{Rotational diagrams}

The rotational diagram, also known as the Boltzmann diagram approach, is a 
rather common methodology used to derive the rotational temperature and column 
density of a given species. The method is described in many works \citep[see, 
for example, ][]{gol99} and assumes a population of the levels characterized by 
\trot, under LTE conditions.

By taking logarithms to Eq.~B.8, and after some algebra, we obtain:

\begin{equation}
\log\left[N(J,K)/g_{JK}\right]=-\frac{\log e}{T_\mathrm{rot}}\ \left(E_{JK}/\mathrm{k}\right) + 
\log\left[N(\mathrm{NH}_3)/Q(T_\mathrm{rot})\right]\,.
\end{equation}

We see in the above equation that the energy of the upper levels are linearly 
related to the logarithms of the column densities. 

If three or more lines of a given species are measured, we can define the 
abscissa $x$ as $\left(E_{JK}/\mathrm{k}\right)$, and the ordinate $y$ as 
$\log\left[N(J,K)/g_{JK}\right]$. $x$ and $y$ are related by the usual linear 
equation $y=a\,x+b$, and we can therefore find the least square regression 
line. \trot\ and $N($\nh$)$ can be obtained from the slope $a$ and the constant 
term $b$ by:

\begin{equation}
T_\mathrm{rot}=-\log e/a
\end{equation}

\noindent
and

\begin{equation}
N(\mathrm{NH}_3)=10^b\ Q(T_\mathrm{rot})\,.
\end{equation}

\end{appendix}

\end{document}